\documentclass{aastex63}

\usepackage{natbib}
\usepackage{xcolor}
 
\newcommand\species[2]{#1 {\sc #2}}

\def\eg{\mbox{e.g.}}
\def\etal{\mbox{\rm et al.}}
\def\teff{\mbox{$T_{\rm eff}$}}
\def\logg{\mbox{log~{\it g}}}
\def\kmsec{\mbox{km~s$^{\rm -1}$}}

\def\vmicro{$V_{mic}$}
\def\carbiso{{$^{12}$C/$^{13}$C}}

\shorttitle{HPF Chemical Compositions}
\shortauthors{Sneden et al.}

\begin{document}

\title{CHEMICAL COMPOSITIONS OF RED GIANT STARS FROM HABITABLE ZONE 
       PLANET FINDER SPECTROSCOPY}

\correspondingauthor{Christopher Sneden}
\email{chris@verdi.as.utexas.edu}

\author{Christopher Sneden}
\affiliation{Department of Astronomy and McDonald Observatory,
             The University of Texas, Austin, TX 78712, USA;
             chris@verdi.as.utexas.edu}
\author{Melike Af\c{s}ar}
\affiliation{Department of Astronomy and Space Sciences,
             Ege University, 35100 Bornova, \.{I}zmir, Turkey;
             melike.afsar@gmail.com}
\affiliation{Department of Astronomy and McDonald Observatory,
             The University of Texas, Austin, TX 78712, USA;
             chris@verdi.as.utexas.edu}
\author{Zeynep Bozkurt}
\affiliation{Department of Astronomy and Space Sciences,
             Ege University, 35100 Bornova, \.{I}zmir, Turkey;
             zeynep.bozkurt@ege.edu.tr}
\author{Gamze B\"{o}cek Topcu}
\affiliation{Department of Astronomy and Space Sciences,
             Ege University, 35100 Bornova, \.{I}zmir, Turkey;
             gamzebocek@gmail.com}
\author{Sergen {\" O}zdemir}
\affiliation{Department of Astronomy and Space Sciences,
             Ege University, 35100 Bornova, \.{I}zmir, Turkey;
             sergenozdemir58@gmail.com}
\author{Gregory R. Zeimann}
\affiliation{Hobby Eberly Telescope,
             The University of Texas, Austin, TX 78712, USA;
             gregz@astro.as.utexas.edu}
\author{Cynthia S. Froning}
\affiliation{Department of Astronomy and McDonald Observatory,
             The University of Texas, Austin, TX 78712, USA;
             cfroning@astro.as.utexas.edu}
\author{Suvrath Mahadevan}
\affiliation{Department of Astronomy \& Astrophysics, The Pennsylvania 
             State University, University Park, PA 16803, USA;
             suvrath@astro.psu.edu}
\affiliation{Center for Exoplanets and Habitable Worlds, The Pennsylvania 
             State University, University Park, PA 16803, USA}
\author{Joe P. Ninan}
\affiliation{Department of Astronomy \& Astrophysics, The Pennsylvania 
             State University, University Park, PA 16803, USA;
             jpn23@astro.psu.edu}
\affiliation{Center for Exoplanets and Habitable Worlds, The Pennsylvania 
             State University, University Park, PA 16803, USA}
\author{Chad F. Bender}
\affiliation{Department of Astronomy and Steward Observatory,
             University of Azizona, Tucson, AZ 85721, USA;
             bender@email.arizona.edu}
\author{Ryan Terrien}
\affiliation{Department of Physics and Astronomy, Carleton College,
             Northfield, MN 55057, USA;
             rterrien@carleton.edu}
\author{Lawrence W. Ramsey}
\affiliation{Department of Astronomy \& Astrophysics, The Pennsylvania 
             State University, University Park, PA 16803, USA;
             lwr@astro.psu.edu}
\affiliation{Center for Exoplanets and Habitable Worlds, The Pennsylvania 
             State University, University Park, PA 16803, USA}
\author{Karin Lind}
\affiliation{Department of Astronomy, Stockholm University, AlbaNova 
             University Centre, SE-106 91 Stockholm, Sweden:
             karin.lind@astro.su.se}
\author{Gregory N. Mace}
\affiliation{Department of Astronomy and McDonald Observatory,
             The University of Texas, Austin, TX 78712, USA;
             gmace@astro.as.utexas.edu}
\author{Kyle F. Kaplan}
\affiliation{SOFIA Science Center, NASA Ames Research Center, Mail Stop 
             N232$-$12 P.O. Box 1, Moffett Field, CA 94036, USA;
             kkaplan@usra.edu}
\author{Hwihyun Kim}
\affiliation{Gemini Observatory, Casilla 603, La Serena, Chile;
             hkim@gemini.edu}
\author{Keith Hawkins}
\affiliation{Department of Astronomy and McDonald Observatory,
             The University of Texas, Austin, TX 78712, USA;  
             keithhawkins@utexas.edu}                    
\author{Brendan P. Bowler}
\affiliation{Department of Astronomy and McDonald Observatory,
             The University of Texas, Austin, TX 78712, USA;  
             bpbowler@astro.as.utexas.edu}

\begin{abstract}

We have used the Habitable Zone Planet Finder (HPF) to gather high resolution, 
high signal-to-noise near-infrared spectra of 13 field red horizontal-branch 
(RHB) stars, one open-cluster giant, and one very metal-poor halo red giant.
The HPF spectra cover the 0.81$-$1.28 \micron\ wavelength range of the 
$zyJ$ bands, filling in the gap between the optical (0.4$-$1.0~\micron) and 
infrared (1.5$-$2.4~\micron) spectra already available for the program stars.
We derive abundances of 17 species from LTE-based 
computations involving equivalent widths and spectrum syntheses, and 
estimate abundance corrections 
for the species that are most affected by departures from LTE in RHB stars.  
Generally good agreement is found between HPF-based metallicities and abundance
ratios and those from the optical and infrared spectral regions.
Light element transitions dominate the HPF spectra of these red giants, and 
HPF data can be used to derive abundances from species with poor or no 
representation in optical spectra (\eg, \species{C}{i}, \species{P}{i},
\species{S}{i}, \species{K}{i}).
Attention is drawn to the HPF abundances in two field solar-metallicity 
RHB stars of special interest:  one with 
an extreme carbon isotope ratio,
and one with a rare very large lithium content.
The latter star is unique in our sample by exhibiting very strong 
\species{He}{i} 10830~\AA\ absorption.
The abundances of the open cluster giant 
concur with those derived from other wavelength regions.
Detections of \species{C}{i} and \species{S}{i} in HD~122563 are 
reported, yielding the lowest metallicity determination of [S/Fe] from
more than one multiplet.

\end{abstract}


\section{INTRODUCTION}\label{intro}

Red horizontal branch (RHB) stars (also called 
``secondary red clump''; e.g., \citealt{girardi98}, \citealt{girardi99}, 
\citealt{ruiz18}) are mostly known for their prominent locations on the 
Hertzsprung-Russell (HR) diagrams of globular clusters. 
They are evolved stars with double energy sources, burning helium in the core 
and hydrogen burning in the shell. 
RHB stars are not easy to identify among field stars, but their cooler 
counterparts, the red clump (RC) stars, are relatively more numerous and
have narrowly constrained temperatures and gravities, making them stand out
in color-magnitude diagrams.
RHB and RC stars have small luminosity ranges, and thus can serve as standard 
candles.

Accurate chemical compositions of RHB stars along with 
robust statistics of their occurrence in disk and halo populations
can enhance our knowledge of stellar and Galactic evolution.
RHBs are important astrophysical tools that can be used to study stellar 
densities, kinematics, and chemical abundances across the Galactic disk, and 
allow us to reach out greater distances than the dwarfs or other regular 
giants would provide (e.g. \citealt{girardi16}). 
\cite{afsar18a} (here after Af\c{s}ar18a) have studied the detailed chemical 
compositions and kinematics of a sample of 340 candidate field RHB stars 
in the optical spectral region (3400$-$10900~\AA). 
Their chemo-kinematical analysis of these candidates revealed the presence of 
more than 150 field RHB and RC stars, located both in the thin and thick disk 
of the Galaxy. 
Later, \cite{afsar18b} (hereafter Af\c{s}ar18b) investigated three RHB stars 
in detail using high-resolution H- and K-band (1.48$-$2.48 \micron) spectra
obtained with the Immersion Grating Infrared Spectrograph (IGRINS).
In that work they argued that in general the IR and optical
abundances agree well, and for some elements (\eg, Mg, Si) the IR abundances 
are more trustworthy than the optical ones.
Additionally, that work highlighted the abundances of several elements 
(\eg, P, S, K) with prominent transitions in the IR whose optical lines are
too weak to be detected or too strong to yield reliable abundances.

Our work investigates some selected RHB stars from Af\c{s}ar2018a,b in
the relatively under-studied $zyJ$ wavelength region, using high-resolution
spectra obtained with the Habitable Zone Planet Finder (HPF) spectrograph.
We derive metallicities and abundances of 13 RHB stars.
We also report on our analyses of the well-known very metal-poor halo red 
giant, HD~122563 (\citealt{afsar16} and references therein, hereafter 
Af\c{s}ar16), and of one red giant member from open cluster NGC 6940 (MMU 105;
\citealt{bocek16}, \citealt{bocek19}, here after BT16, and BT19).
These two stars will be considered separately from the discussion of the RHB
stars.

In \S\ref{obsred} we introduce the stellar sample, 
describe HPF, and outline the observations and reductions.
Atomic and molecular line selection, equivalent width measurements, synthetic 
spectrum computations, and abundance derivations are given in \S\ref{analyses}.
We consider four stars with noteworthy abundance signatures in 
\S\ref{goodstars}, and conclude with our summary in \S\ref{discussion}.

\vspace*{0.2in}
\section{OBSERVATIONS AND REDUCTIONS}\label{obsred}

The red giants investigated in this HPF spectroscopic study 
(Table~\ref{tab-log}) have been selected from our previous studies
(Af\c{s}ar16,18a,b, BT16,19).
See those papers for detailed descriptions of the optical and IGRINS observations
of our program stars.
In general the HPF $zyJ$ wavelength domain has been relatively neglected in 
stellar abundance studies.
In order to sample spectra of red giants in this region, we have 
selected stars with various chemical composition characteristics. 
The following sub-sections focus on the details of HPF observations.

\subsection{HPF Spectra}\label{hpfobs}

HPF is a high$-$resolution 
($\lambda/\Delta\lambda$~$\equiv$~$R$~$\sim$~55,000) near$-$infrared ($zyJ$ 
photometric bands, 8100$-$12800~\AA) spectrograph on 
the 10m Hobby Eberly Telescope (HET) at McDonald Observatory
\citep{mahadevan12,mahadevan14}.\footnote{
https://hpf.psu.edu/}
The main HPF goal is to search for exoplanets of cool M dwarf stars by 
detecting the orbital reflex motions of the parent stars.
To achieve high velocity precision ($\sim$1.5~m~s$^{-1}$), the entire HPF 
instrument is encased in a vacuum chamber cooled to 180 Kelvin, yielding 
temperature stability of $\sim$~1~mK.
The instrument has input fiber scrambling \citep{roy14}, and precise 
wavelength characterization with 
a laser frequency comb \citep{halverson14,metcalf19}.
At present HPF has generated studies of exoplanet systems and precision radial 
velocity techniques, but it has not been routinely applied to investigation of
the atmospheres of
stars warmer than spectral type M.

High-resolution, high signal-to-noise ($S/N >$ 100) HPF 
spectra of 15 target stars were gathered over several months using
the fully queue-scheduled HET \citep{shetrone07}.
Table~\ref{tab-log} gives basic data for these stars, and for warm, rapidly
rotating, mostly featureless stars used as telluric divisors.
In the $zyJ$ spectral region the telluric molecular blockage ranges 
from near zero to almost total, varying significantly with wavelength.
The dominant telluric contaminators are H$_2$O bands, and their contamination 
can change substantially on timescales of minutes, usually requiring a 
hot-star observation to accompany each target star.
Some mitigation is provided by the fixed altitude of the HET, which on
stable nights can lead to successful use of divisor star spectra acquired
some time before or after the target spectra.

HPF observations of target and telluric warm stars were reduced using the 
HPF pipeline code $Goldilocks$.\footnote{
https://github.com/grzeimann/Goldilocks\_Documentation}
This package processed the raw HPF data by removing bias noise, correcting for 
nonlinearity, masking cosmic rays, and calculating the slope/flux and 
variance image using the algorithms from the pyhxrg module in the tool 
$HxRGproc$ \citep{ninan18}.  
The pipeline also processed bright lamp exposures to combine into a master 
frame to model the trace of the calibration, sky, and science 
fibers as well as their respective fiber profiles.   
The profiles were used as weights in an optimal extraction algorithm 
(\citealt{horne86}) to produce 1D spectra and their associated propagated 
errors for all 28 orders. 
The master frame was also used to correct flat fielding features.

$Goldilocks$ uses a single wavelength solution for all observations from the 
well-calibrated NIR Laser Frequency Comb.  
HPF's wavelength solution is extremely stable but does drift on the order of 
$\sim$10-20~m~s$^{-1}$ throughout the night with long term drifts over months 
on the same scale.  
This amounts to $\sim$1/100th of a pixel. 
Although $Goldilocks$ provides the drift correction in the observation file 
headers, it is only important for precision radial velocity studies; 
we ignored this very small effect in our analysis.

For each stellar source, the science fiber extractions from $Goldilocks$ 
includes light from the target as well as OH emission from the night sky. 
The sky fiber, located 50.1\arcsec away, contains only night sky emission.
In most cases, the relative flux ratio between the sky emission and stellar 
continuum (at the wavelengths of interest) is less than 5\%, 
and the sky emission has almost no effect on the target spectrum. 
In those situations we used the spectra of targets as 
extracted from the science fibers without attempting sky subtraction.
For a few small spectral regions in which sky subtraction was critical to 
accurate representation of the stellar spectra, we performed the sky 
subtraction manually using the pipeline-extracted star and sky spectra.

After these reduction procedures, we performed continuum 
normalization of the target spectra via the \textit{continuum} task 
of IRAF\footnote{
http://iraf.noao.edu/}, 
and we used the \textit{telluric} task
to divide out the telluric contamination from the science target spectra. 
Later, individual orders were merged into a single continuous spectrum with
the \textit{scombine} task.

The $Goldilocks$ pipeline produces rest wavelength solutions that 
are accurate to levels far beyond what are needed for our study.
To measure the radial velocities ($RVs$) of our stellar targets
we made use of \textit{rv} package in IRAF. 
First we created synthetic spectra (as outlined in \S\ref{lines}) adopting 
similar \teff, \logg, and [Fe/H] values to our program stars. 
Then we used \textit{rvcorrect} for the heliocentric corrections, and 
measured the $RV$ values with the cross-correlation task \textit{fxcor} 
\citep{fitzpatrick93}.
We only worked on the wavelength regions that are not affected (or only 
affected minimally) by the atmospheric telluric lines: 8420$-$8520 \AA, 
8540$-$8640 \AA, 8660$-$8760 \AA, 10275$-$10400 \AA\, and 10450$-$10575 \AA. 
We list the $RV$s from HPF and their standard deviations ($\sigma$) 
for each star in Table~\ref{tab-vel}. 
The $\sigma$ values were calculated by taking the average of $RVs$ measured 
from individual regions listed above. Previous $RV$ measurements of the same
targets (Af\c{s}ar18a and $Gaia$ DR2) are also listed in Table~\ref{tab-vel} 
for comparison purposes.

The internal $RV$ accuracies (line-to-line scatters) for individual stars 
are comparable:
$\langle\sigma_{HPF}\rangle$~=~0.18~\kmsec\ and 
$\langle\sigma_{opt}\rangle$~=~0.19~\kmsec,
and the velocity scales agree:  
$\langle RV_{opt} $-$ RV_{HPF}\rangle$~= $+$0.45~$\pm$~0.51~\kmsec\ 
($\sigma$~=~1.96~\kmsec).
Our program stars also have $RV$s in the Gaia Data Release 2 Catalog 
\citep{arenou18},and we list them also in Table~\ref{tab-vel}.
The agreement betwen HPF and Gaia is very good:
$\langle RV_{opt} $-$ RV_{Gaia}\rangle$~= $-$0.27~$\pm$~0.08~\kmsec\
($\sigma$~=~0.31~\kmsec).

\subsection{Other Spectra}\label{otherobs}

Our program stars have all had previous extensive analyses from
optical region spectra.
As described in detail by Af\c{s}ar18a, the aHB high-resolution spectra were 
obtained with the Robert G. Tull Cross-Dispersed Echelle spectrograph 
\citep{tull95} on the McDonald Observatory 2.7m Harlan J. Smith telescope. 
The instrumental setup yielded a spectral resolving power of 
$R$~=~60,000 with effective wavelength coverage for
the RHB observations of 4100$-$9000~\AA.
Redward of 5000~\AA\ the $S/N$ values always exceeded 100.
The Tull Spectrograph was also used to obtain a very high $S/N$ spectrum
of the very metal-poor HD~122563.
The optical spectrum for open cluster giant NGC~6940 MMU 105 was acquired
with the High Resolution Spectrograph \citep{tull98} of the Hobby-Eberly 
Telescope at McDonald Observatory.  
It was configured to cover the spectral range 5100$-$6900~\AA, with the
same resolving power as our 2.7m Tull Spectrograph data.

A few of our HPF stars have also been observed at high resolution in the 
infrared $H$ and $K$ bands with the Immersion Grating Infrared Spectrometer 
(IGRINS; \citealt{yuk10,park14}) when this portable instrument was located 
at the McDonald Observatory 2.7m telescope.
The IGRINS spectra for these bright RHB stars covered a wavelength range 
14,800$-$24,800~\AA, had resolving power $R$~$\simeq$~45,000, and
achieved $S/N$~$\geq$~150.
Our metallicity and abundance ratio analysis was reported by Af\c{s}ar18b.

\vspace*{0.2in}
\section{ABUNDANCE ANALYSES}\label{analyses}

\subsection{Line Lists}\label{lines}

\begin{figure}
\epsscale{0.90}
\plotone{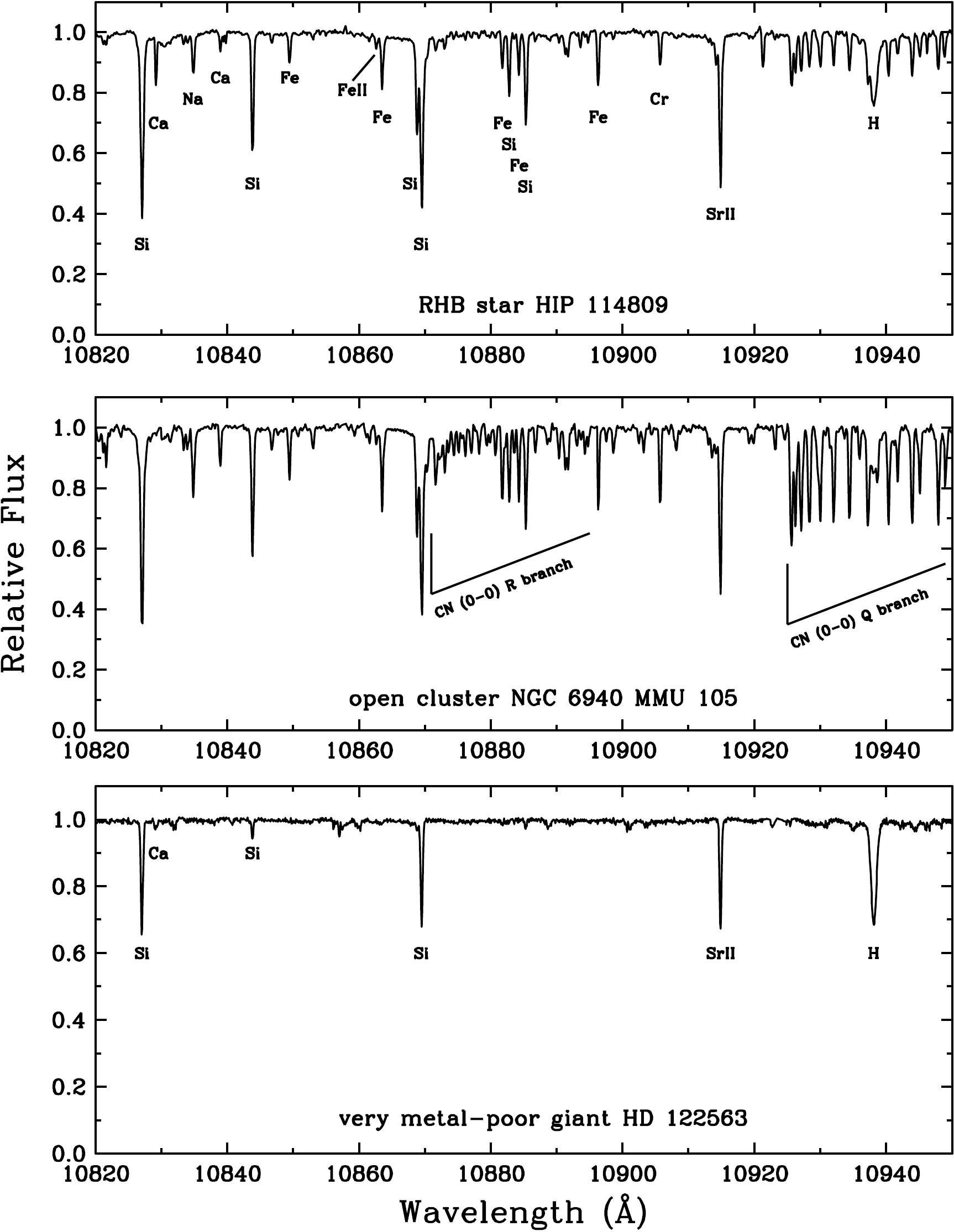}
\caption{\label{fig1}
\footnotesize
   Spectra in an HPF echelle order that has only 
   moderate telluric line contamination (removed in the reduction process 
   described in \S\ref{hpfobs}).
   The top panel shows a typical RHB program star, the middle panel has the
   single cooler$-$temperature open cluster red clump giant, and the bottom
   panel has the very metal$-$poor field giant HD~122563.
   In the top panel some prominent atomic lines are labeled.
   All identified lines except \species{Fe}{ii} and \species{Sr}{ii} are
   neutral-species transitions.
   The CN (0$-$0) R-branch bandhead is also indicated.
}
\end{figure}

In this exploration of HPF spectra of red giants we first sought to 
identify all useful atomic and molecular transitions for abundances
analysis.
HPF spectra cover the $zyJ$ wavelength range 8170$-$12780~\AA\ 
(0.82$-$1.28~\micron).
In RHB stars this spectral region contains many atomic lines, nearly all of
which arise from light and Fe$-$group elements.
In Figure~\ref{fig1} we show one of the HPF 
spectral orders that has several strong neutral$-$species lines of Na, 
Si, Ca, Fe, and Cr, one of the \species{H}{i} Paschen series lines, 
\species{Fe}{ii} 10862.65~\AA, and \species{Sr}{ii} 10914.87~\AA.
The \species{Sr}{ii} line and its multiplet partners at 10036.66 and 
10327.31~\AA are $zyJ$ region rarities:  ionized-species neutron-capture 
element ($Z$~$>$~30) transitions.
Neutron$-$capture elemental transitions are rarely strong enough in our stars
for detection in this wavelength domain.
A few Fe$-$group ionized species lines can be seen in red giant stars, but
only one \species{Fe}{ii} multiplet is strong enough for abundance analysis in
our program stars.

There are also many molecular features in the $zyJ$ region: 
OH ground state $^1\Sigma^+$, CO ground state X$^2\Pi$ rovibrational, 
and CN A$^2\Pi$$-$X$^2\Sigma ^+$ red system transitions.
But in general only CN lines have large absorption strengths.
They can significantly contaminate many otherwise useful atomic transitions.
In practice this problem forces examination of each potential atomic line
for the presence of blends.

With potential line contamination issues, we chose candidate atomic features 
for abundance analysis by comparing HPF data in each spectral order with 
synthetic spectra for RHB star HIP~114809.
Detailed matching of synthetic and observed spectra was not sought at this
point.
For all computations of synthetic spectra and predicted single-line equivalent
widths ($EW$s) in this paper we used the current version of the LTE 
plane-parallel line analysis code MOOG \citep{sneden73}\footnote{
Available at http://www.as.utexas.edu/$\sim$chris/moog.html}.
The input atomic and molecular line lists were generated by the
auxiliary code \textit{linemake}\footnote{
Available at https://github.com/vmplacco/linemake}.
We adopted the model atmosphere for HIP~114809 derived for the RHB survey
by Af\c{s}ar18a, and whose parameters are given in Table~\ref{tab-models}.
By this process we identified useful transitions of 16 elements; 14 of these
have only neutral-species lines.  
A single \species{Ti}{ii} line was detected, 
along with the very strong \species{Sr}{ii} lines mentioned above.

In Table~\ref{tab-lines} we list the atomic lines chosen for this study.  
A fundamental limitation of abundance analyses in the $zyJ$ spectral region
is the lack of substantial sets of reliable transition probabilities
based on laboratory spectroscopy.
For this initial HPF abundance study of RHB stars we decided to avoid deriving
empirical log($gf$) values from the observed solar spectrum, as their values
are difficult to disentangle from choices of solar model atmosphere, line 
damping parameters, and solar microturbulent velocity.
The Wisconsin lab atomic physics group has published transition 
probabilities in the HPF domain for \species{Ti}{i} \citep{lawler13}, 
and \species{Co}{i} \citep{lawler15}.
For \species{Fe}{i} a collaboration between investigators at Imperial 
College London Blackett Laboratory and the Wisconsin group 
\citep{obrian91,ruffoni14,denhartog14,belmonte17} has yielded an extensive set 
of reliable $gf$ values ranging from the $UV$ to the near-$IR$.
We used only lines from those studies in our Ti, Fe, and Co analyses.
This ensured $gf$ source consistency for these elements in HPF and optical
spectra, at the expense of ignoring potentially useful transitions in
the HPF region.

Of the remaining 13 species, there are 8 with detected lines in our HPF data
that have transition probabilities given in the NIST atomic line database
\citep{kramida19}\footnote{
https://physics.nist.gov/PhysRefData/ASD/lines\_form.html}.
The NIST atomic line catalog is a critical compilation, and they provide
transition probability accuracy assessments.
For our lines these quality estimates range from ``A+'' (uncertainties less 
than 2\%) to ``D'' (uncertainties less than 50\%).  
We list the NIST line qualities in Table~\ref{tab-lines}.
We have included all promising lines without discrimination among NIST 
quality values, but caution should be observed for those lines with
ratings of ``C'' and lower.

Finally, detected HPF transitions  of \species{Ca}{i}, \species{Cr}{i}, 
\species{Mn}{i}, and \species{Ni}{i} lack recent lab transition work and they
do not appear in the NIST database.
For these lines we adopted the $gf$$-$values given in the 
\cite{kurucz11,kurucz18}\footnote{
http://kurucz.harvard.edu/linelists.html}
semiempirical line compendium.
Decades of work by Kurucz have yielded a database of millions of transitions.
Pragmatic mixing of results from laboratory and theoretical studies along with
fresh computations has produced good matches between stellar observed
and predicted spectra, but individual transition probabilities have 
widely varying accuracies; this should be kept in mind here.

\species{S}{i} deserves a more expanded description.
There are several lines in the $zyJ$ 
spectral domain, and they generally have $gf$-values available in the
NIST database.
Table~1 of \cite{spite11} lists the most promising transitions in the optical
and near-IR regions. 
In our HPF spectra of RHB stars we have all of these lines except the 
optical triplet at 6757~\AA\ (see \citealt{costasilva20} for a large survey 
based on those lines).
Detections in our spectra include the ground-state [\species{S}{i}] line
at 10821.11~\AA.  
\cite{caffau07b} have explored 
the use of this transition in solar-type stars.  
It is an attractive alternative to the high-excitation 
($\chi$~$\gtrsim$ 6.5~eV) \species{S}{i} lines that are normally studied,
because it has little \teff\ sensitivity and is essentially formed in LTE.
However, there are some practical problems with this line for our abundance
analyses: (a) the line is very weak; (b) it has an un-identified blend in
its red wing 
(see Figure~1 in \citeauthor{caffau07b});
and (c) it is located only 2~\AA\ from the edge of its HPF spectral order.
For the present HPF exploration we drop the 10821~\AA\ line, but it
should considered more carefully in future papers.

The \species{S}{i} triplet at 9212.9, 9228.1, and
9237.5~\AA\ is strong and has been used in several S abundance studies
(\eg, \citealt{spite11}, \citealt{koch11}).
In our HPF spectra these lines are useful, but with two cautions.
First, they are somewhat saturated in our RHB stars, and the derived S 
abundances have some dependence on assumed microturbulent velocity \vmicro.
Second, the spectral region surrounding this \species{S}{i} has a large 
amount of telluric contamination, which is a potential problem with HPF
located at McDonald Observatory. 
The papers cited here were based on data obtained at Chilean observatories,
with much less water vapor contaminating the stellar spectra.

The \species{S}{i} triplet near 10457~\AA\ is about
half as strong as the 9200~\AA\ one, and for metal-rich giants such as
our RHBs these lines are easy to analyze.
Finally, \cite{ryde19} has identified a line at 10635~\AA\  as a \species{S}{i}
transition, particularly for evolved stars with \teff~$\gtrsim$~5500~K (see
their Figure~1).
Preliminary syntheses confirmed its suitability for our study.

In the end, the preliminary synthetic spectrum 
computations outlined here yielded about 100 atomic lines worthy of further 
abundance analysis.  
Most of these were judged relatively uncontaminated by known atomic and 
molecular blends, and thus could be safely analyzed from their $EW$
measurements.
About 20 of the lines were either somewhat blended, or very weak, 
or have hyperfine structure.
For these we performed synthetic spectrum calculations.

\subsection{Equivalent Widths}\label{ewmeasures}

For the majority of  transitions we derived abundances from $EW$ computations.
We measured $EW$s with the specialized software package $SPECTRE$
\cite{fitzpatrick87}\footnote{
Available at http://www.as.utexas.edu/$\sim$chris/spectre.html}.
Interactive fits were made for all lines, with most of them modeled with
Gaussian approximations.
For the strongest lines, we used empirically-determined Voigt profiles.
We also set local continua interactively in this procedure, since the IRAF 
$continuum$ task produce good general fits to each HPF order blaze function
but are inadequate for the small spectral intervals surrounding each 
measured line.

\begin{figure}
\epsscale{0.75}
\plotone{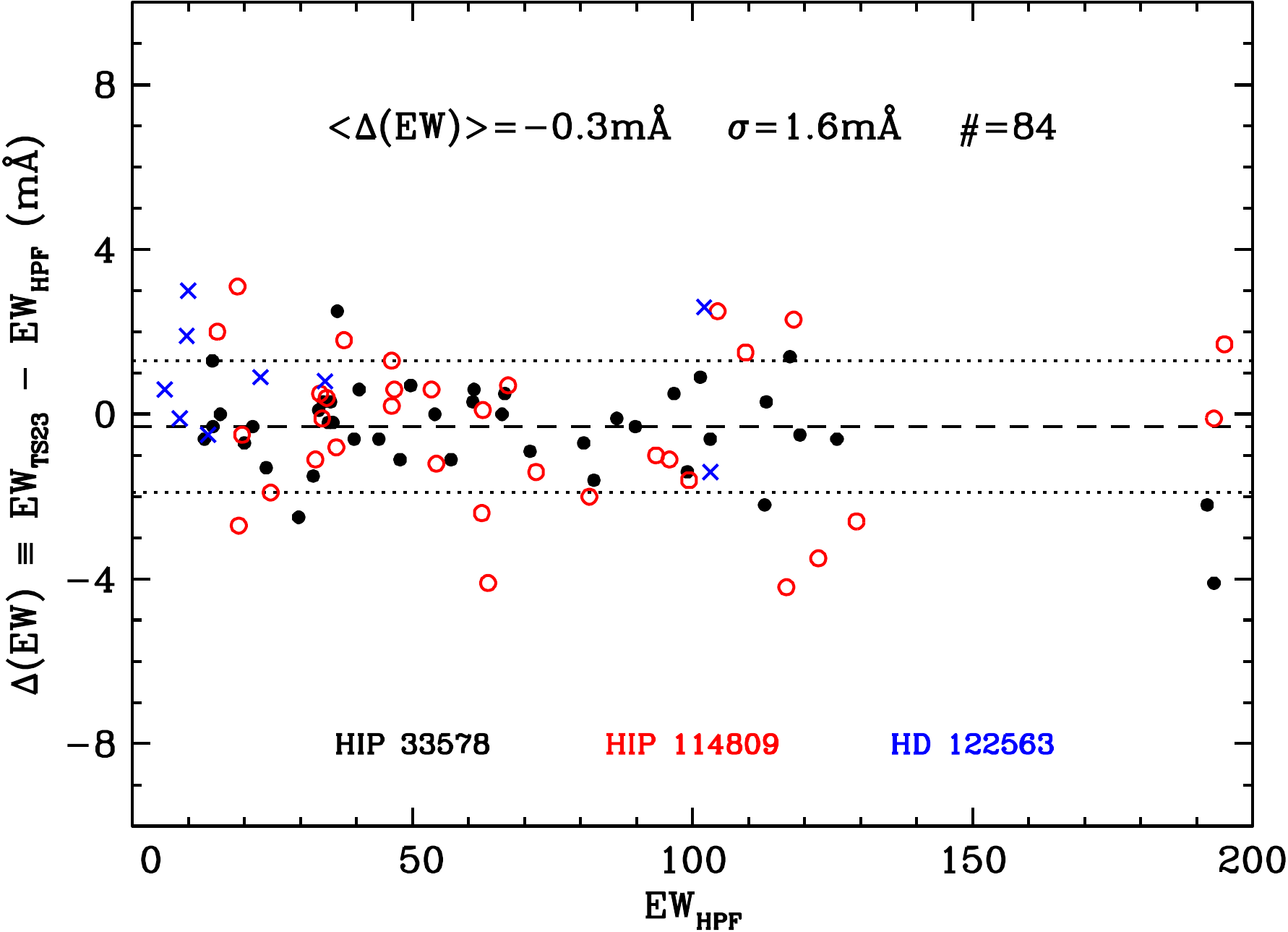}
\caption{\label{fig2}
\footnotesize
   Comparison of equivalent widths measured with the McDonald Observatory
   TS23 optical echelle spectrograph \citep{tull95} and the HPF.
   The dashed line represents the mean difference, 
   $\langle\Delta(EW)\rangle$ = $-$0.3~m\AA, and the dotted lines represent the 
   standard deviation of individual measurements $\sigma$~= 1.6~m\AA.
   Three stars, identified in the figure legend, were used for this test.
   See \ref{ewmeasures} for details.
}
\end{figure}

The short wavelength limit for HPF is $\sim$8080~\AA.
This overlaps with the long wavelength limit of the optical spectrograph
TS23, $\sim$8900~\AA.\footnote{
The HPF low wavelength limit is defined by the spectrograph setup, whose
disperser positions cannot be altered.
For TS23 the cross-disperser tilt can be changed to put very long wavelength
echelle orders on its 
CCD detector, but (a) the detector quantum efficiency 
strongly declines beyond 9000\AA; (b) the detector suffers increasing amounts
of fringing toward 10000~\AA; and (c) telluric H$_{2}$O absorption lines 
are extremely strong in the $\sim$8850$-$9850~\AA\ wavelength region.}
In the parts of this wavelength domain that are relatively free from 
telluric absorption ($\sim$8300$-$8900~\AA), we measured $EW$s for RHB 
stars HIP~33578, HIP~114809,
and metal-poor HD~122563 using spectra from both instruments.
In Figure~\ref{fig2} we show the $EW$ comparisons.
The agreement between HPF and TS23 values is at the level of our ability
to reliably make $EW$ measurements.
Decisions on continuum placement and line profile fits cause 
$EW$ measurement scatter at the level indicated by the line-to-line scatter.
For the strongest lines, assumptions of appropriate line profile
shape (Gaussian or Voigt) also can affect the $EW$s, but the fractional
$EW$ differences in weak and strong lines are similar.
The two spectrographs yield essentially identical spectra of red giant stars.

\subsection{Adopted Model Atmospheres}\label{mods}

The model atmospheric parameters of our science targets have been adopted from 
our previous efforts (Af\c{s}ar16, BT16, Af\c{s}ar18a), in which we describe 
the derivation of model atmospheres in detail. 
In summary here, we used a semi-automated version of the MOOG code introduced 
in \S\ref{lines}.
This driver software begins with information from photometry and spectra 
line-depth ratios to generate initial model parameter estimates, and then 
operates iteratively using both neutral and ionized species of Fe and Ti 
lines to calculate the best model atmospheric parameters. 
They are listed in Table~\ref{tab-models}.
Typical parameter uncertainties are 150~K for \teff, 0.25 for \logg\ and 
\vmicro, and 0.1 for [Fe/H]; see Af\c{s}ar18a \S 6.3 for a discussion of 
these values.
Optical atomic and molecular line lists 
that we used to derive atmospheric parameters and individual 
element abundances were adopted from \cite{bocek15} and BT16.

\subsection{Derived Abundances}\label{abunds}

We determined the abundances of most elements by matching their equivalent
widths with computed ones, using the stellar atmosphere models 
(Table~\ref{tab-models}) and their line parameters (Table~\ref{tab-lines})
described above.
Iterations on the output abundances and re-examinations of the spectra resulted
in identification and elimination of aberrant lines in individual stars.
In Table~\ref{tab-lineabs} we give line-by-line 
abundances in log~$\epsilon$ units.\footnote{
For elements A and B,
[A/B] = log $(N_{A}/N_{B})_{\star}$ -- log $(N_{A}/N_{B})_{\sun}$
and log $\epsilon$(A) = log $(N_{A}/N_{H})$ + 12.0 .
Also, metallicity will be taken to be the [Fe/H] value.} 
Program star metallicities [Fe/H] computed from the \species{Fe}{i}  and
\species{Fe}{ii} lines are listed in Table~\ref{tab-models}.
To compute the [Fe/H] values and subsequent abundance ratios [X/Fe] we have 
assumed the solar abundance set recommended by \cite{asplund09}.

\begin{figure}
\epsscale{0.75}
\plotone{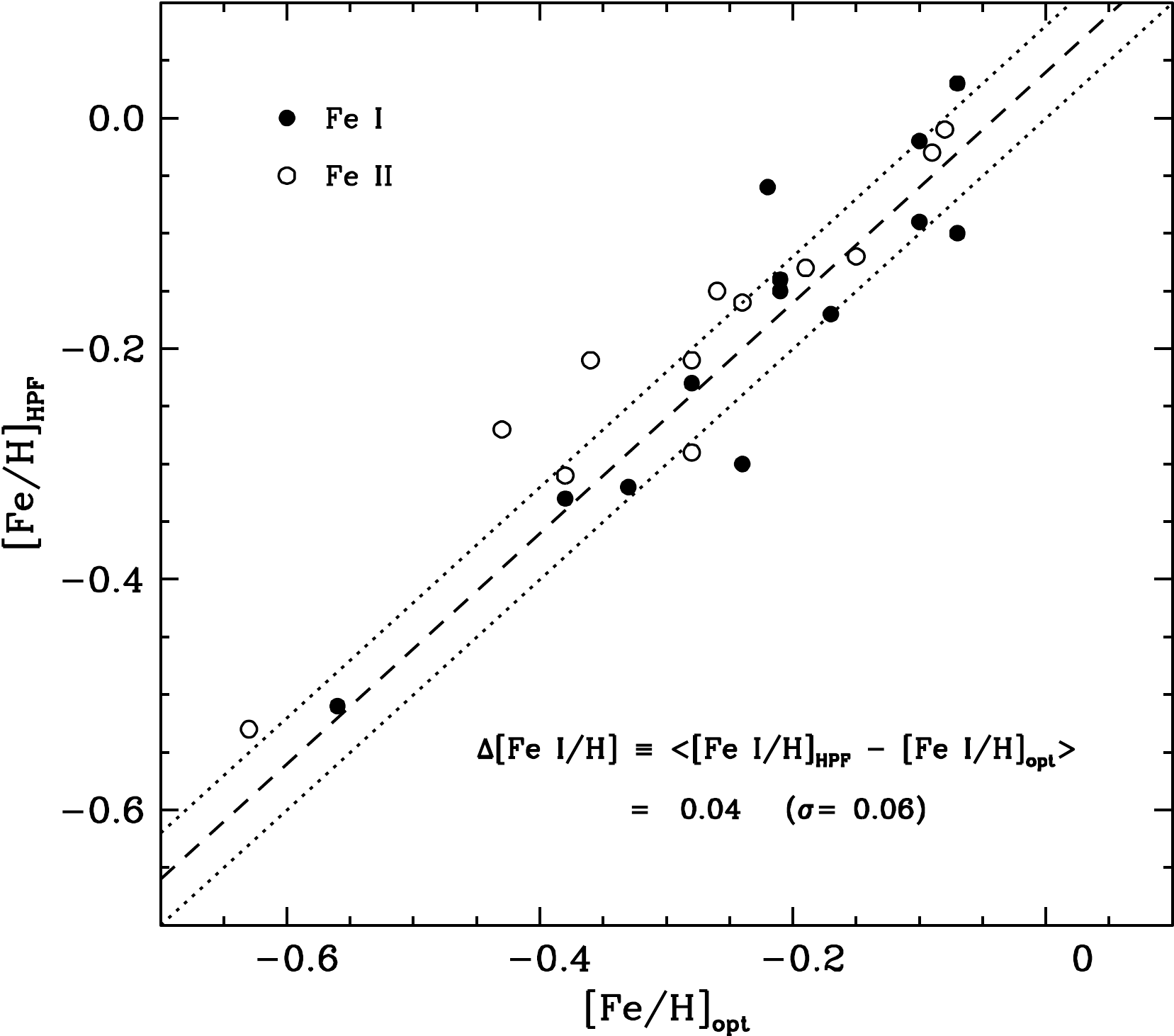}
\caption{\label{fig3}
\footnotesize
   Comparison of \species{Fe}{i} and \species{Fe}{ii} 
   metallicities with those from optical spectra Af\c{s}ar18a after 
   adjustment to the solar abundances of \cite{asplund09}.
   The dashed line represents the mean offset between the two \species{Fe}{i}
   sets, and the dotted lines represent the standard deviation.
}
\end{figure}

HPF metallicities of RHB stars are in reasonable accord 
with those derived from optical data (Af\c{s}ar18a). 
In Figure~\ref{fig3} we illustrate the comparisons.
The mean difference for \species{Fe}{i} between the two metallicity sets, 
defined as $\Delta$[\species{Fe}{i}/H]~$\equiv$ 
$\langle$[\species{Fe}{i}/H]$_{\rm HPF}$ $-$ 
[\species{Fe}{i}/H]$_{\rm opt}\rangle$ = 0.04 ($\sigma$~= 0.06).
Abundances from both spectral regions are based on the same Wisconsin/London
group transition probability sources.
The comparison for \species{Fe}{ii} shows a larger offset,
$\Delta$[\species{Fe}{ii}/H]~=~0.08 ($\sigma$~= 0.05).
All of the optical and HPF $gf$-values for \species{Fe}{ii} are taken from
the NIST database, and nearly all of them originate with a critical compilation
by \cite{raassen98}, who considered calculated and experimental transition
probabilities in their study.  
The internal consistency of the \species{Fe}{ii} HPF lines is good, since
the typical line-to-line abundance scatter in this species is small, 
$\langle\sigma\rangle$~$\simeq$~0.05 (Table~\ref{tab-models}).
However, the relationship between transition data for these lines and those
used in optical analyses should be re-examined if larger surveys are
undertaken in the future. 

\begin{figure}                                                
\epsscale{0.90}                                               
\plotone{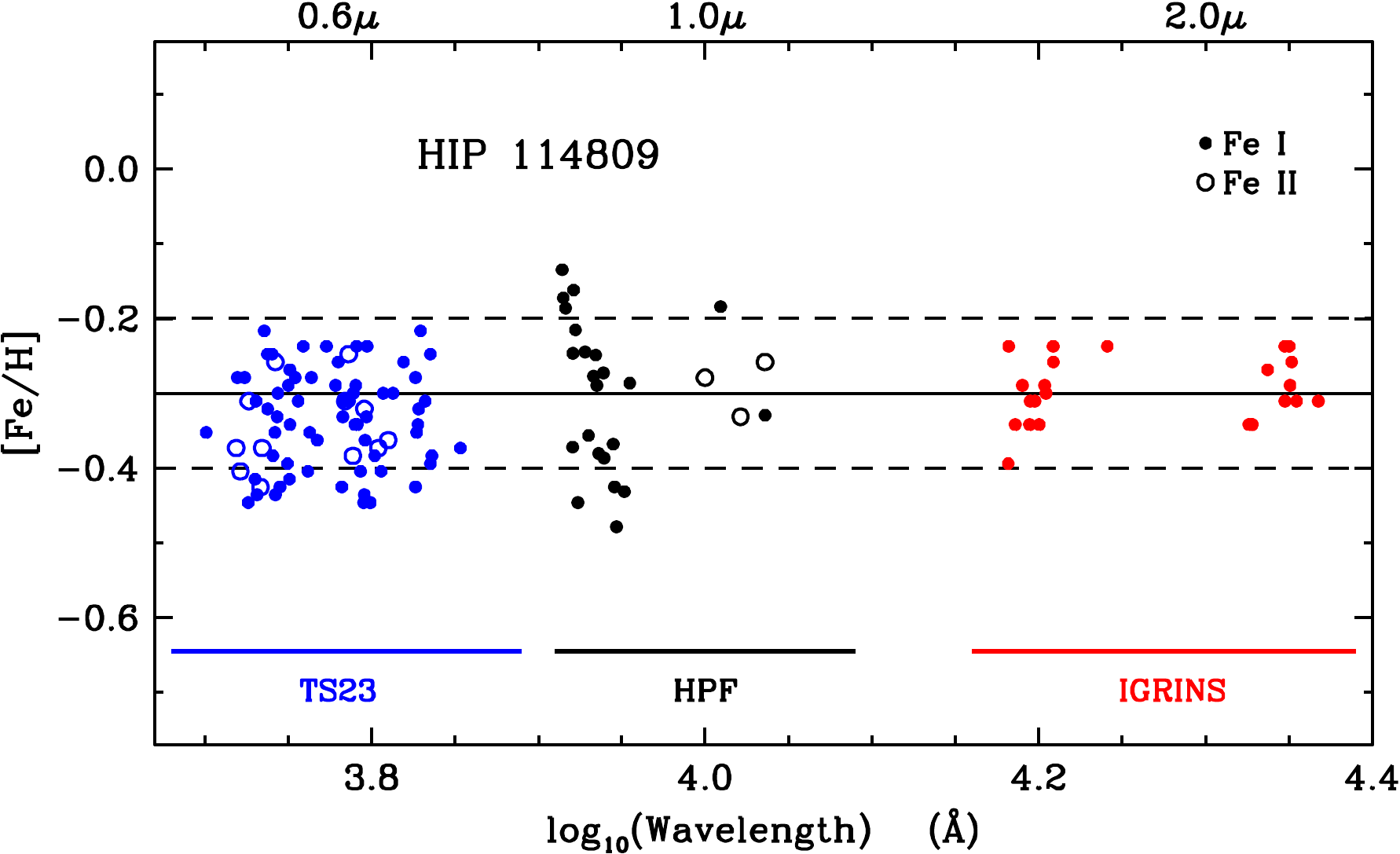}                                          
\caption{\label{fig4}                                       
\footnotesize                                                 
   Abundances of Fe lines plotted as a function of wavelength.  
   The wavelength axis is plotted logarithmically due to its large range.
   Results from optical TS23, HPF, and IGRINS spectra are shown with 
   different colors.
   The lines denote statistics of the HPF \species{Fe}{i} results; the solid 
   line is for $\langle$[\species{Fe}{i}/H]$\rangle_{\rm HPF}$, and the two
   dashed lines represent the 1-$\sigma$ standard deviation for the HPF
   abundances.
}                                                             
\end{figure}

For a few stars in our sample we have optical, 
HPF $zyJ$, and IGRINS $HK$ spectroscopic analyses, extending from 
$\sim$5000~\AA\ to almost 24,000~\AA.  
In Figure~\ref{fig4} we plot the three abundance sets for HIP~114809. 
Inspection of this figure reveals agreement among the abundances 
(Table~\ref{tab-models}) within the line-to-line scatter.
Mean values for \species{Fe}{i} and \species{Fe}{ii} range over
$\langle$[Fe/H]$_{\rm opt}\rangle$ = $-$0.38 to $-$0.31, and 
and $\langle$[\species{Fe}{i}/H]$_{\rm IGRINS}\rangle$ = $-$0.32, 
$\sigma$~=~0.04 (Table 6 in Af\c{s}ar18b).
The mean Fe abundances have similar scatter in other program stars.

The optical and HPF \species{Fe}{i} abundances have been derived with 
laboratory transition probabilities 
\citep{obrian91,denhartog14,ruffoni14,belmonte17}.
Unfortunately, there are very few recent laboratory log($gf$) values for
this species (and for most other species as well) for wavelengths beyond
10,000~\AA.
Therefore the [Fe/H]$_{\rm IGRINS}$ abundances in Af\c{s}ar18b were 
determined with transition probabilities derived from reverse solar analyses.
Additionally, the formal $\sigma$-values and the appearance of 
Figure~\ref{fig4} show that the line-to-line scatter is larger in the 
HPF abundances than in the optical ones.
At least two factors may account for this effect.
First, on average the near-$IR$ \species{Fe}{i} lines are much weaker than 
optical and $UV$ ones.
They come from relatively weak branches of multiplets, and therefore they
often have larger transition probability uncertainties than those stronger
lines at short wavelengths.\footnote{
This transition probability problem is even more acute in other species,
\eg, \species{Fe}{ii}, which has a rich spectrum of lines in the $UV$
($\lambda$~$<$~3000~\AA), but only relatively weak transitions at longer
wavelengths \citep{denhartog19}.
Only a few potentially useful (but ultimately discarded) \species{Fe}{ii}
lines were found in our HPF spectra of RHB stars, and there are none in
the IGRINS wavelength region.}
Second, all of the HPF \species{Fe}{ii} lines occur in the range
8200$-$9010~\AA.  
In this region the CN contamination can be severe.
We have derived \species{Fe}{ii} abundances with full synthetic spectra, 
and discarded lines that appear to be severely blended. 
Probably future abundance surveys can successfully extract reliable 
abundances from $EW$ analyses of our chosen \species{Fe}{ii} lines, but 
caution is warranted.

\subsection{Abundance Ratios}\label{xtofe}

\begin{figure}                                                
\epsscale{1.00}                                               
\plotone{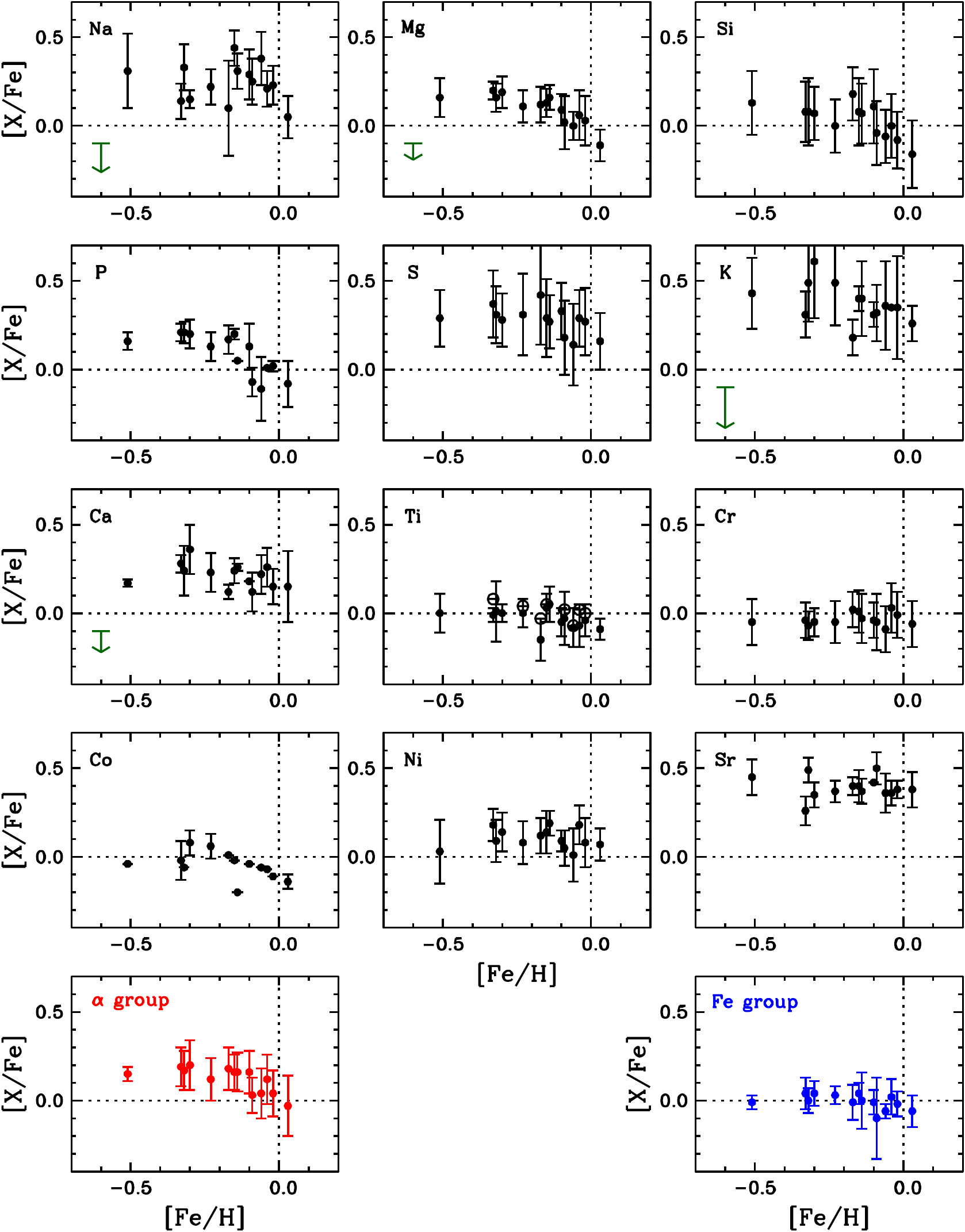}                                          
\caption{\label{fig5}                                       
\footnotesize                                                 
   Abundance ratios [X/Fe] plotted as functions of the 
   [Fe/H] metallicity for most elements in this study.
   The solar values of these quantities, [Fe/H]~$\equiv$ 0.0 and 
   [X/Fe]~$\equiv$ 0.0, are indicated with dotted lines.
   In the panel for Ti (middle panel of the figure), the abundances derived 
   from \species{Ti}{ii} lines are depicted with open symbols.
   In the Fe-group (bottom middle) panel, the points are colored blue to
   emphasize that they are simple means of Ti, Cr, Mn, Co and Ni abundances.
   In the $\alpha$-group (bottom right) panel, the points have red colors;
   they are simple means of Mg, Si, S, and Ca abundances.
}                                                             
\end{figure}

Mean abundance ratios [X/Fe] are presented in Table~\ref{tab-meanabs},
and in Figure~\ref{fig5} we show the trends of [X/Fe] with [Fe/H]
metallicity.
Carbon abundances are not plotted in this figure; they will be discussed
in \S\ref{cno}.
A few general remarks are in order here.
First, we re-emphasize that we have enforced single-source transition 
probability choices for individual species, but there is $gf$ heterogeneity 
from species to species; see \S\ref{lines} for details.
Second, we have relied on standard LTE analyses. 
Our estimates of non-LTE corrections will be presented in \S\ref{nltecalc}.
Third, the total metallicity domain is small, 
$-$0.5~$\lesssim$~[Fe/H] $\lesssim$~0.0, including only thin-disk and
some thick-disk stars (see Af\c{s}ar18a for stellar population 
assessments of RHB stars).
Meaningful discussion of Galactic trends awaits analysis of a sample over
a much wider metallicity range.

With these cautions in mind, it is clear in Figure~\ref{fig5} that 
Fe-group elements Ti, Cr, Mn, Co and Ni all have solar abundance ratios, 
[X/Fe]~$\simeq$~0, from HPF spectra as they do in abundance surveys in other 
spectral regions.
Forming averages of these elements for each star, and defining
[Fe-group/Fe]~$\equiv$ 
[$\langle$\species{Ti}{i},\species{Ti}{ii},\species{Cr}{i},\species{Mn}{i},\species{Co}{i},\species{Ni}{i}/Fe$\rangle$],
for the whole Fe group, the mean for the 13 star RHB set is 
$\langle$[Fe-group/Fe]$\rangle$~=~$-$0.01 ($\sigma$~=~0.04).
The Fe-group abundance means for each star are shown in the bottom middle
panel of Figure~\ref{fig5}, showing no trend with metallicity, as expected
from previous Galactic disk abundance surveys (\eg, 
\citealt{reddy03,reddy06,bensby14}, Af\c{s}ar18a)

The $\alpha$ elements Mg, Si, S, and Ca show a small trend of increasing
[X/Fe] values with decreasing [Fe/H].
Defining [$\alpha$/Fe]~$\equiv$ [$\langle$Mg, Si, S, Ca/Fe$\rangle$],
we plot the mean $\alpha$ values for the RHB stars in the bottom right 
panel of Figure~\ref{fig5}.
At [Fe/H]~$\simeq$~0.0, [$\alpha$/Fe]~$\simeq$~0.0, and for stars with
[Fe/H]~$\lesssim$~$-$0.2, [$\alpha$/Fe]~$\simeq$~$+$0.2, again in agreement
with the trends found in the survey cited above.

\subsection{Non-Local Thermodynamic Equilibrium Computations}\label{nltecalc}

Local thermodynamic equilibrium (LTE) can be an adequate 
approximation for abundance computations in red giants for ionized majority
species, \eg, the lanthanide ions.
However, almost all of our HPF transitions arise from neutral species.
Most of their parent elements are heavily ionized in line-forming atmospheric 
levels, and their neutral transitions in the HPF domain arise from 
high-excitation levels (averaging $\sim$4.5~eV in our line list),
and many of them have deep line cores formed in shallow layers of the 
atmosphere where photon losses are strong.
We expect that some non-LTE corrections are necessary to 
produce accurate abundances for most of our HPF transitions.

New detailed non-LTE abundance corrections for 13 elements 
have been published by \cite{amarsi20}, including Na, Mg, Si, K, and Ca.
The abundance corrections discussed in that study were limited to the optical 
transitions of most interest to the GALAH survey \citep{desilva15}.
For this paper we have estimated the non-LTE abundance shifts for as many
of our HPF transitions as possible using the departure coefficients by
\citeauthor{amarsi20}

The LTE and non-LTE synthetic spectra were computed
with $PySME$, the python version of the spectroscopic analysis code 
$SME$ \citep{piskunov17}.
The computations here are initial estimates of the magnitudes of non-LTE 
effects for each HPF line in a typical RGB star.
A more detailed investigation will be carried out in the future.

For now, we feel more confident with the general magnitude 
of the non-LTE corrections, and so in Table~\ref{tab-nltecorr} we present mean 
values, standard deviations, and number of lines involved in the calculations 
for six neutral species for a typical RHB star and for HD~122563 (to be 
discussed in \S\ref{hd122563}).
The entries in this table suggest that for HPF lines in RHB stars the typical
non-LTE abundance correction is $\langle\Delta_{corr}\rangle$~$\simeq$~$-$0.15
with a line-to-line scatter $\sigma$~$\simeq$~0.10.
In Figure~\ref{fig5} we illustrate the $\langle\Delta_{corr}\rangle$ values
with green arrows in the panels for Na, Mg, K, and Ca.
It is apparent that in each of these cases the estimated non-LTE shift
significantly reduces the apparent LTE-based overabundances.

\subsection{The CNO Group}\label{cno}

Our RHB stars have few CNO strong abundance indicators in the $zyJ$ spectral 
range.  
There are some easily-identified \species{C}{i} lines, but this species has 
played only a very minor role in the extensive literature on C abundances 
in evolved cool stars.
Additionally, there are no strong \species{O}{i} or OH lines in this wavelength
domain, thwarting any attempt to derive O abundances.
Two prominent CN red-system (0$-$0) bands are seen in the metal-rich 
stars of Figure~\ref{fig1}: the R-branch head at $\simeq$10871~\AA\ 
and the Q-branch head at $\simeq$10925~\AA. 
But N abundances derived from these bands are dependent on derived or assumed
C and O abundances.

Even with these limitations some headway on CNO can be made with HPF spectra.
In Table~\ref{tab-licno} we summarize optical and HPF abundances for the 
RHB stars and NGC 6940 MMU 105.
The data from optical spectra will be reported by Bozkurt et al. (2021, 
in preparation).
That work will give details of the abundance derivations and will include
discussion of uncertainties.
The final [C$_{mean}$/Fe] values are based on multiple C indicators:
the CH G-band in the 4280$-$4330~\AA\ range, and C$_2$ Swan bands with
bandheads near 5160 and 5630~\AA.
Abundances from  \species{C}{i} are determined and tabulated by Bozkurt \etal\
but they do not participate in the C abundance means.
The optical N abundances have been determined using C$_{mean}$ abundances
derived earlier in concert with the O abundances 
(from the [\species{O}{i}] line).
Carbon isotopic ratios come mostly from the 8004~\AA\ CN feature that is
the basis for most \carbiso\ estimates in red giants.

Abundances from HPF \species{C}{i} transitions in Table~\ref{tab-licno} are
repeated from Table~\ref{tab-meanabs}.  
Mean values for the stellar sample have not been computed because star-to-star 
abundance differences for C, N, and \carbiso\ are natural in evolved giants.
The three sets of C abundances in Table~\ref{tab-licno} are offset from 
each other but they do correlate well:
$\langle$[C$_{C I,opt}$/Fe]$-$[C$_{mean,opt}/Fe]\rangle$ = $+$0.10, 
$\sigma$ = 0.08;
$\langle$[C$_{C I,HPF}$/Fe]$-$[C$_{C I,opt}/Fe]\rangle$ = $+$0.19, 
$\sigma$ = 0.08; and
$\langle$[C$_{C I,HPF}$/Fe]$-$[C$_{mean,opt}/Fe]\rangle$ = $+$0.30, 
$\sigma$ = 0.12.
The \species{C}{i} HPF lines clearly yield aberrantly high abundances compared
to the other values.
The uncertainties in the mean differences are consistent with uncertainties
in the observations and abundance computations involving \species{C}{i} lines.
This issue should be investigated in the future, with attention to accuracy and 
consistency of transition probability sources, stellar atmosphere parameter
dependences, and non-LTE sensitivities of these high-excitation 
($\chi$ $\gtrsim$ 7.5 eV) transitions.

Abundances of N were determined via synthetic/observed spectrum matches
of the CN (0-0) Q-branch wavelength regions, with the redward-degrading
bandhead culminating at 10925~\AA.
The synthetic spectra were generated assuming the
optical O and C$_{mean}$ abundances of Table~\ref{tab-meanabs}.
Formally the optical and HPF N abundances agree very well:
$\langle$[N$_{CN,HPF}$/Fe]$-$[N$_{CN,opt}/Fe]\rangle$ = $+$0.01, 
$\sigma$ = 0.05.
However, the CN transition data used here all come from one recent extensive
laboratory study \citep{brooke14}, and \cite{sneden14} have shown that
various CN bands from blue through the near-$IR$ yield internally consistent
abundance results.
The red system (0-0) bands in the HPF region are much stronger than the
(2-0) bands near 8000~\AA\ that have dominated CN studies in the optical
spectral region.
Probably this will enable CN detection in warmer RHB stars than those
included in this study.

Finally, we attempted to estimate \carbiso\ ratios from the $^{13}$CN (0-0)
Q-branch bandhead that occurs at 10923~\AA, almost 3~\AA\ blueward of the
$^{12}$CN bandhead.
However, in our RHB stars the $^{13}$CN lines are very weak, and this 
small spectral region is contaminated by telluric lines that are difficult 
to remove accurately.
Therefore Table~\ref{tab-licno} lists only nine isotopic ratios, and a couple of
these have large estimated uncertainties.
For most stars the best conclusion is that the HPF \carbiso\ values are
consistent with, but probably inferior to those derived from the optical
(2-0) transitions.

\section{Stars of Special Interest}\label{goodstars}

Af\c{s}ar18a identified about 150 true field RHB giants out of their original
candidate sample of $\sim$340 stars.
Our RHB targets to observe with HPF included two that Af\c{s}ar18b 
analyzed with IGRINS H- and K-band spectra, 12 chosen at random from the 
Af\c{s}ar18a list, and two that seemed to be worth special attention.
Here we discuss these two RHB field stars, the one open cluster giant,
and metal-poor HD~122563.

\subsection{HIP~33578: Undiluted Interior CN$-$Cycle Products on the Surface}\label{hip033578}

\begin{figure}
\epsscale{1.00}
\plotone{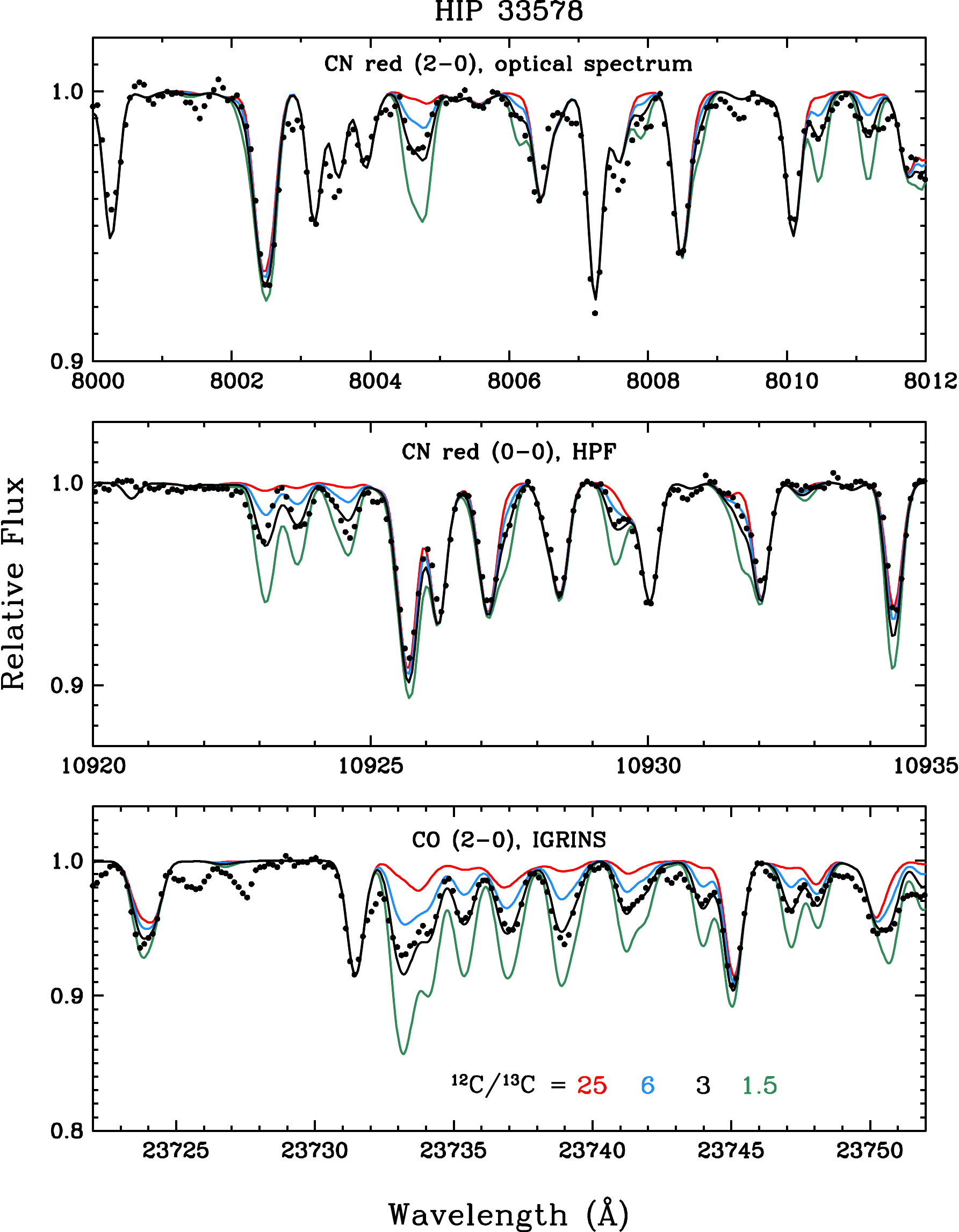}
\caption{\label{fig6}
\footnotesize
   Prominent $^{13}$C molecular features in three wavelength domains of
   HIP~33578.
   The dots represent the observed spectra, and the lines represent synthetic
   spectra with \carbiso\ values defined in the figure legend by the line 
   colors.
}
\end{figure}

HIP~33578 is an RHB field star that was an unremarkable member of the 
Af\c{s}ar18a survey.
But that study concentrated on kinematics, Fe metallicities, and limited
Fe-group and $\alpha$ element abundance ratios of its RHB sample.
Bozkurt \etal\ (in preparation) will provide much more extensive abundance
information on the LiCNO elements that are sensitive to stellar interior
fusion cycles and envelope mixing, along with results many other element
groups.
In this work we have noticed that HIP~33578 is one of the rare metal-rich 
field giants with a very low carbon isotopic ratio.  
Fortunately we were also able to obtain an IGRINS H- and K-band spectrum
of this star.
The optical, IGRINS, and now HPF spectra present a solid observational case 
for the extreme \carbiso\ value, as we illustrate in Figure~\ref{fig6}.
For all three wavelength regions, the C abundance is the same while
the N abundance has been allowed to vary $<$0.1~dex to best match the observed
$^{12}$CN and $^{12}$CO in each small spectral interval.
The best estimate from synthetic/observed spectrum matches from each of these
molecular bands is \carbiso~=~3~$\pm$~1.
This ratio at the surface of HIP~33578 is lower than, but consistent within 
the observational and theoretical uncertainties, of the interior CN-cycle 
\carbiso\ value presented first discussed in detail by \cite{caughlan62}.

Further support comes from the C and N abundances.  
From Table~\ref{tab-licno} the mean values for RHB stars excluding HIP~33578 
are $\langle$[C/Fe]$\rangle$~=~$-$0.36 ($\sigma$~=~0.14) and 
$\langle$[N/Fe]$\rangle$~=~0.57 ($\sigma$~=~0.14).  
The optical results for HIP~33578 are [C/Fe]~=~$-$0.88 and [N/Fe]~=~0.67 
(in accord with the HPF N abundance).
The 0.5~dex extra depression in C and more modest enhancement of N are 
consistent with the very low \carbiso\ of this star.

The LiCNO abundances and carbon isotopic ratio of HIP~33578 are in the domain 
of the ``weak G-Band'' (wkG) disk giants \citep{sneden78,adamczak13,palacios16}.
For HIP~33578 Bozkurt \etal\ (2021, in preparation), supported by our HPF 
results, have derived 
log~$\epsilon$(Li)~$<$ 0.0, [C/Fe]~= $-$0.7, [N/Fe]~= 0.7, [O/Fe]~= 0.2, 
and \carbiso~= 3, while approximate mean values from 
\cite{adamczak13} and \cite{palacios16} are
$\langle$log~$\epsilon$(Li)$\rangle$~= 3 to 10, 
$\langle$[C/Fe]$\rangle$~$\simeq$ $-$1.4, 
$\langle$[N/Fe]$\rangle$~$\simeq$ 1.1,
$\langle$[O/Fe]$\rangle$~= 0.0, and 
$\langle$\carbiso$\rangle$~$\simeq$ 3 to 10.
HIP~33578 has evidence of less C$\rightarrow$N conversion.
This star's Li abundance is very much smaller than the typical wkG
giant, but the samples of both \cite{adamczak13} and \cite{palacios16} have
stars with no obvious Li enhancement.
Our single star cannot provide much new insight into the evolutionary history
of this small red giant subclass.
However, the RHB status of HIP~33578 is shared by many wkG stars.
From its \teff~= 5118~K  and its photometry and parallax given in 
\cite{afsar18a} we suggest that log(L/L$_\odot$~$\simeq$ 2.5, in the middle
of the set of warmer wkG stars shown in the HR diagrams of
\cite{palacios16} (see their Figures~7 and 8).
We tentatively assign HIP~33578 to the wkG subclass.

\subsection{HIP~99789: a rare Red Horizontal Branch Star with Extremely High Lithium}\label{hip099789}

\begin{figure}
\epsscale{1.00}
\plotone{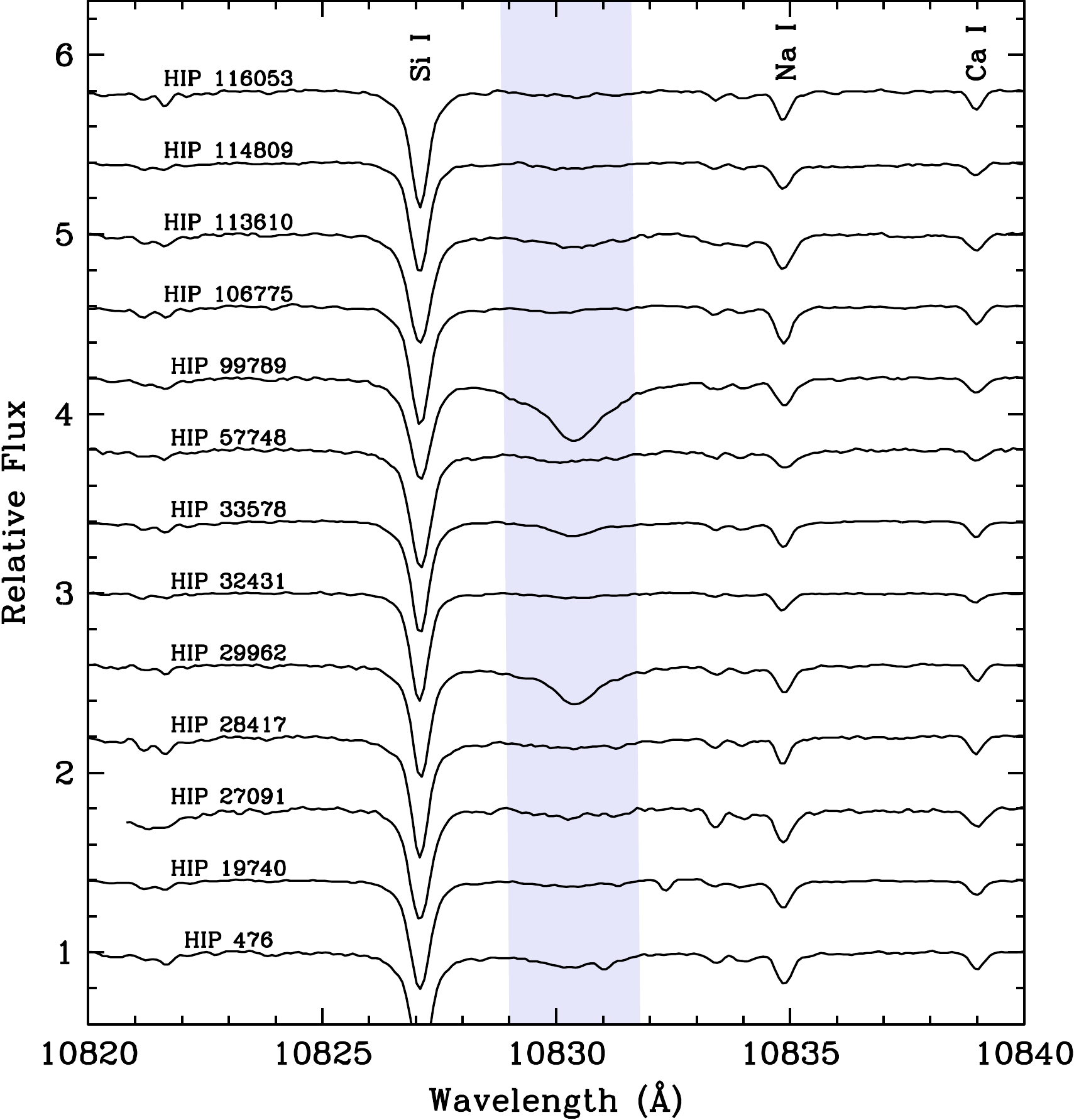}
\caption{\label{fig7}
\footnotesize
   Spectra of the RHB stars surrounding the \species{He}{i} 10830~\AA\ line.  
   The spectra are ordered by star name.
   The lavender-shaded wavelength area marks the approximate wavelength
   domain of the broad \species{He}{i} transition.
   The relative flux scale for HIP 476 is correct, and vertical offsets
   of 0.4 have been added in succession to the spectra of the other 14 RHBs.
   Three prominent photospheric absorption lines have been identified,
   and there are smaller atomic and CN lines that can bee seen in the plot.
}
\end{figure}

In contrast to HIP~33578, the RHB star HIP~99789 has ordinary CNO abundances
and \carbiso\ for our RHB sample (Table~\ref{tab-licno})  but an extremely 
high Li content:  log~$\epsilon$(Li)~$\simeq$ 2.6, close to the interstellar
medium value.
No other abundance anomaly is evident in this star.
However, we have detected an unusually strong \species{He}{i} absorption 
line at 10830 \AA.
In Figure~\ref{fig7} we show a montage of our RHB spectra in the wavelength 
region surrounding the 10830 \AA\ line.
This transition arises from a very high-excitation (19.8~eV) metastable
level of \species{He}{i}. 
It has been used extensively in past studies to trace solar/stellar 
chromospheric activity and wind outflows, and has gained recent attention as
a potential indicator of outflows from exoplanet atmospheres (\eg,
\citealt{ninan20}).
Inspection of Figure~\ref{fig7} suggests that more than half of our RHB sample,
8 out of 13 stars, have extremely little or no detectable \species{He}{i} 
absorption at 10830 \AA.
Another three stars, HIP~476, HIP~33578, and HIP113610, have modest
\species{He}{i} lines.
But HIP~29962 and HIP~99789 have strong 10830~\AA\ absorptions, suggesting
substantial chromospheric activity in their atmospheres.
We cannot identify any other distinguishing spectroscopic signatures in
HIP~99789, but the possible connection between high Li abundance and 
\species{He}{i} activity
is intriguing and should be pursued with a
dedicated set of HPF observations of a larger sample of high-Li stars.

Bozkurt \etal\ (in preparation) will show that HIP~99789 is the most 
Li-rich member of the Af\c{s}ar18a sample.
However, its appearance among the RHBs may match a recent discovery about
evolved stars with high Li.
\cite{singh19} have determined Li abundances in red giants whose evolutionary
status can be assigned to either the first-ascent RG or the red clump
on the basis of their asteroseismological properties 
determined from the Kepler satellite \citep{mathur17}.
Significant Li abundances can only be found in the He core-burning red
clump stars (see their Figure~4).
The standard red clump is cooler than the RHB; the highest \teff\ reported by 
\citeauthor{mathur17} for their Li-rich stars is 4999~K, and almost all other
stars are cooler than 4900~K.
Most of our RHBs have \teff~$\gtrsim$~5100~K.
But for HD~99789 Af\c{s}ar18a derived \teff~=~5054~K and \logg~=~2.41
(Table~\ref{tab-models}).
The temperature is not radically different than those of the 
\citeauthor{mathur17} Li-rich sample, and the gravity is 
in the middle of their \logg\ distribution.
HIP~99789 seems to support their assertion that Li enhancement is
strongly associated with the He-core burning phase of late stellar evolution.

However, many decades ago \cite{alexander67} speculated that red
giants could temporarily enhance their Li contents by ingesting companion
planets in their bloated envelopes.
Some evidence connecting Li-rich giants to planetary systems has accumulated
recently, \eg, \cite{adamow18} and references therein.
HIP~99789 may have one or more stellar or planetary companions, as 
indicated by its abnormally large apparent spatial acceleration.
Close binary companions can influence unresolved spectra by diluting line 
depths of the primary star, adding in new lines, and producing anomalous 
RV offsets.  
None of our targets show signs of being double-line spectroscopic binaries
from their RV cross-correlation functions.  
But low mass ratio or longer-period companions could still be present 
in these systems.  
Recently, \citet{brandt19} cross-matched astrometry from \emph{Hipparcos} and 
\emph{Gaia} DR2 to produce the \emph{Hipparcos-Gaia Catalog of Accelerations} 
(HGCA), which provides systematics-corrected proper motions spanning a 
$\approx$24-year baseline between the two missions.
This catalog provides a convenient way to search for tangential accelerations 
(changes in proper motion) that can be attributed to long-period giant planets,
brown dwarfs, low-mass stars, or white dwarfs depending on the amplitude of 
the acceleration and orbital separation (e.g., \citealt{bowler20}).

Following \citet{brandt19}, we make use of the \emph{Gaia} and 
\emph{Hipparcos-Gaia} scaled positional difference proper motions to 
calculate accelerations in our sample.  
Among the 17 stars we observed, all except for NGC6940 MMU 105 have 
\emph{Hipparcos} astrometry and therefore have entries in the HGCA catalog.  
Four out of these 16 stars have significant astrometric accelerations: 
HIP~29962 (149 $\pm$ 22~m s$^{-1}$ yr$^{-1}$), 
HIP 99789 (670 $\pm$ 19~m s$^{-1}$ yr$^{-1}$), 
HIP 113610 (255 $\pm$ 18~m s$^{-1}$ yr$^{-1}$), and 
HIP 114809 (76 $\pm$ 8~m s$^{-1}$ yr$^{-1}$).  
Interestingly, HIP 29962, HIP 99789, and HIP 114809 also have the largest 
discrepancies ($\approx$4~km s$^{-1}$) between the RVs measured from the 
optical Tull data and our HPF spectra, offering further evidence that these 
stars have companions based on radial accelerations.  
The limited astrometric and RV sampling makes it challenging to constrain 
the nature of the companion, but the amplitude of these accelerations 
suggest that the companions are most likely low-mass stars or perhaps 
white dwarfs, both of which are not expected to meaningfully impact the 
spectra and abundances we derive in this study.
While no solid connection can be argued between the large Li of HIP~99789 and
its abnormally large apparent acceleration, it should be studied in the future.

\subsection{NGC 6940 MMU 105: The First Open Cluster Star With Extremely 
Broad High Resolution Spectral Coverage}\label{n6940mmu105}

The cool red giant MMU 105 is a member of the open cluster NGC~6940, and
we have previously reported on its chemical composition from optical (BT16) and
IGRINS (BT19) spectra. 
The instrumental setups and observations were 
like those discussed in \S\ref{otherobs}.
Model atmosphere parameters, metallicities, and abundance ratios for 12
NGC~6940 giants were determined in similar fashion to that used for the RHB
field stars (Af\c{s}ar18a), with the added advantage of having a reliable 
color-magnitude diagram (CMD) for the cluster.
The optical-region metallicity of MMU 105 
([Fe/H]~= $-$0.15\footnote{
As renormalized to the solar abundances of \cite{asplund09}; see BT19.}
$\sigma$~= 0.07) is somewhat smaller than the cluster mean [Fe/H] = $-$0.02 
($\sigma$~=~0.06).
Note that MMU 105 is the coolest (4765 K) of the 12 RGs.
The IGRINS H- and K-band iron abundance is larger, 
$\langle$[Fe/H]$\rangle$ = $-$0.04;
our new HPF value is more consistent with the optical result.
The Fe-group and $\alpha$ elements in this cluster have the abundances of 
slightly metal-poor disk stars.  
For elements with 21~$\leq$~Z~$\leq$~30, 
$\langle$[Fe-group/Fe]$\rangle$~= 0.01~$\pm$~0.14, 
and for lighter elements, $\langle$[$\alpha$/Fe]$\rangle$~= 0.17~$\pm$~0.07.
The neutron-capture elements (Z~$>$~30) appear to be slightly
overabundant by an average of about 0.2~dex.
From IGRINS data, $\langle$[Fe-group/Fe]$\rangle$ = 0.15~$\pm$~0.07, and 
[$\alpha$/Fe]~= [$\langle$Mg, Si, S, Ca/Fe$\rangle$]~= 0.11~$\pm$~0.08.
The HPF abundances listed in Table~\ref{tab-meanabs} concur with the optical
means for these element groups.

The conclusion of BT16 and BT19 from photometric (CMD) and spectroscopic 
(abundance) evidence is that all 12 RGs including MMU 105 are 
helium-burning clump stars.
The abundance argument begins with carbon isotopic ratios:
\carbiso$_{optical}$ = 15$^{+3}_{-2}$, \carbiso$_{IGRINS}$ = 23~$\pm$~3, 
and now we can add \carbiso$_{HPF}$ = 15$^{+5}_{-2}$.
Both optical and IGRINS data reveal underabundances of C and overabundances 
of N.
The HPF N abundances from the (0-0) bandheads agree well with the previous
values, and the HPF \species{C}{i} abundances are in accord with the optical
and IGRINS abundances after accounting for the systematic \species{C}{i}
offset discussed above.
The isochrones discussed in BT16 and BT19 suggest 
that all NGC~6940 RGs are core 
He-burning red clump stars with mostly undergone canonical first dredge-up.
The HPF C, N, and \carbiso\ values should not be the primary data for
assessing mixing in NGC~6940 MMU 105, but clearly they are in agreement with 
our previous optical and IGRINS studies.

\subsection{HD~122563: The Brightest Very Metal$-$Poor Halo Giant}\label{hd122563}

The red giant HD~122563 (HIP~068594) is the only very low metallicity 
member of the Yale Catalog of Bright Stars \citep{hoffleit95}.
It was first noted as a high proper motion star 
\citep{roman55}, and
was one of the first very metal$-$poor stars subject to detailed abundance
analysis \citep{wallerstein63}.
Among many noteworthy chemical composition features, HD~122563 has a very
large relative oxygen abundance 
([O/Fe]~$\simeq$~$+$0.6; \citealt{lambert74}),
very low carbon isotopic ratio (\carbiso~=~5~$\pm$~2; \citealt{lambert77}),
and weak neutron-capture abundances in an $r$-process pattern
\citep{sneden83,honda06}.

The HD~122563 spectrum portion displayed in Figure~\ref{fig1} has 
identifications of three \species{Si}{i} lines and one each of
\species{H}{i}, \species{Ca}{i}, \species{Sr}{ii}.
Neglecting the \species{H}{i} Paschen line, these five transitions represent
a significant fraction (5/34) of the total number of lines that we were
able to employ in the HD~122563 abundance analysis.
This star simply has a metallicity about 500 times lower than the other
stars of our study; most absorption lines simply fade into undetectability.

In Table~\ref{tab-hd122lines} we list the lines employed in our analysis.
The same EW and synthetic spectrum analyses described earlier were applied
to the HPF spectrum of HD~122563.  
We imposed the same transition probability restrictions for this star
that we used for the other stars of our sample (see \S\ref{lines}).
We adopted the same model atmosphere that was used in the IGRINS spectrum
analysis of \cite{afsar16}.
The modest results are presented in Table~\ref{tab-hd122means}.

The HD~122563 HPF metallicity,  
[\species{Fe}{i}$_{HPF}$/H]~= $-$2.87 ($\sigma$~= 0.07),
is consistent with optical value:
[\species{Fe}{i}$_{opt}$/H]~= $-$2.92 ($\sigma$~= 0.12),
Of the other species entered in Table~\ref{tab-hd122means}, we draw 
attention to \species{Si}{i}, \species{S}{i}, and \species{Sr}{ii}.

\begin{figure}
\epsscale{0.75}
\plotone{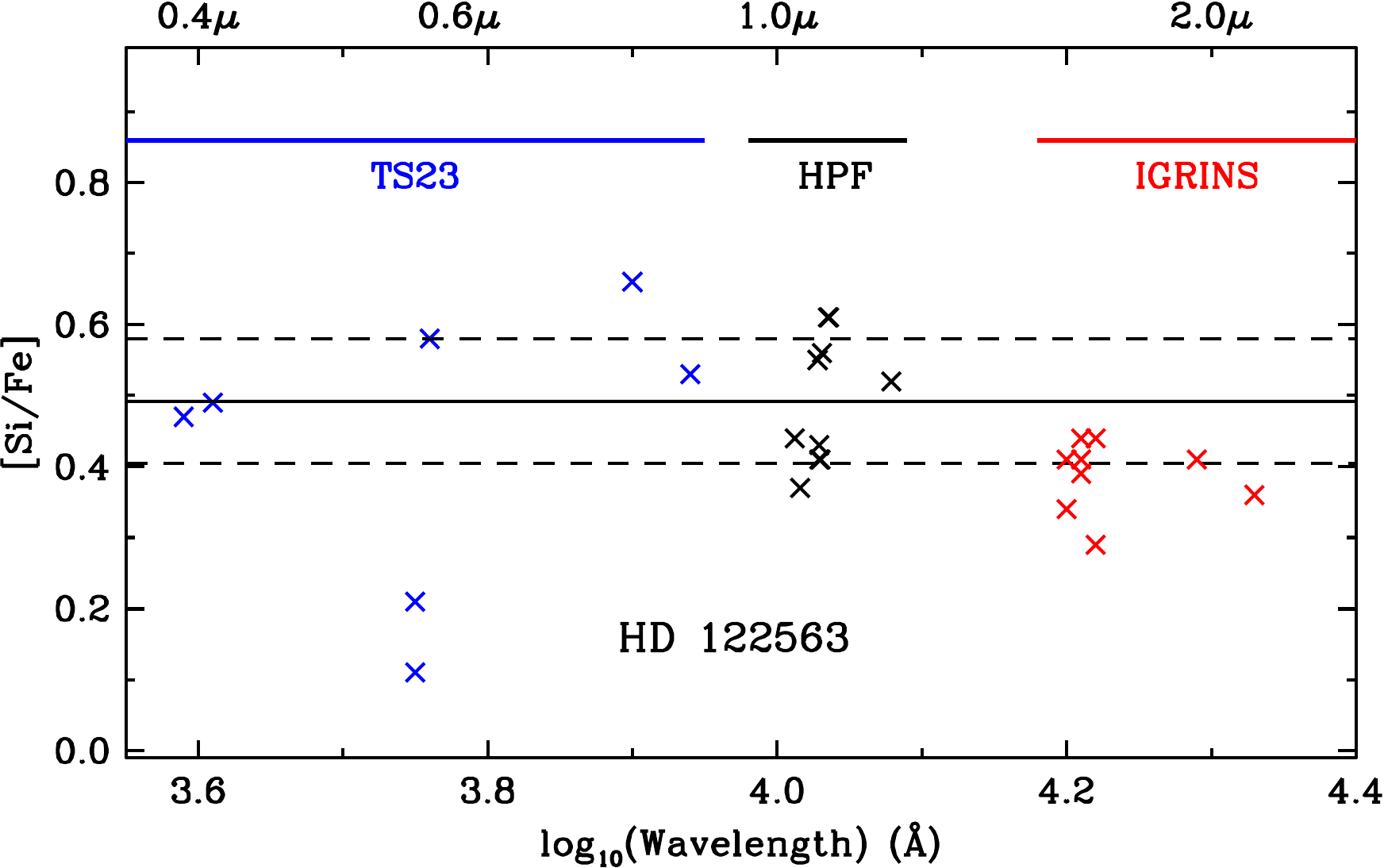}
\caption{\label{fig8}
\footnotesize
   Si abundances from optical (TS23), HPF, and IGRINS data in HD~122563.
   The solid and dashed horizontal lines show the mean and standard
   deviation values for the HPF results.
}
\end{figure}

The abundances from \species{Si}{i} lines in all three spectral domains
are displayed in Figure~\ref{fig8}.
The transition probabilities for all the transitions were taken from
the NIST database.
Si abundances derived from HPF lines perhaps are the most reliable set.
Certainly, as discussed in \S4.2 of \cite{afsar16}, the optical abundances
may be the least trustworthy: the two strong lines in the blue (3905~\AA\
and 4102~\AA) are beset with blending and analytical difficulties, and
the yellow/red lines are all very weak 
(log~$RW$~$\equiv$ log($EW$/$\lambda$)~$\lesssim$ $-$5.6).
The $zyJ$ \species{Si}{i} lines in the HPF spectrum yield a mean Si abundance 
about 0.1~dex higher than do the lines in the IGRINS H- and K-band region.
Resolution of this small discrepancy should begin with a comprehensive
new transition probability study.

\begin{figure}
\epsscale{0.60}
\plotone{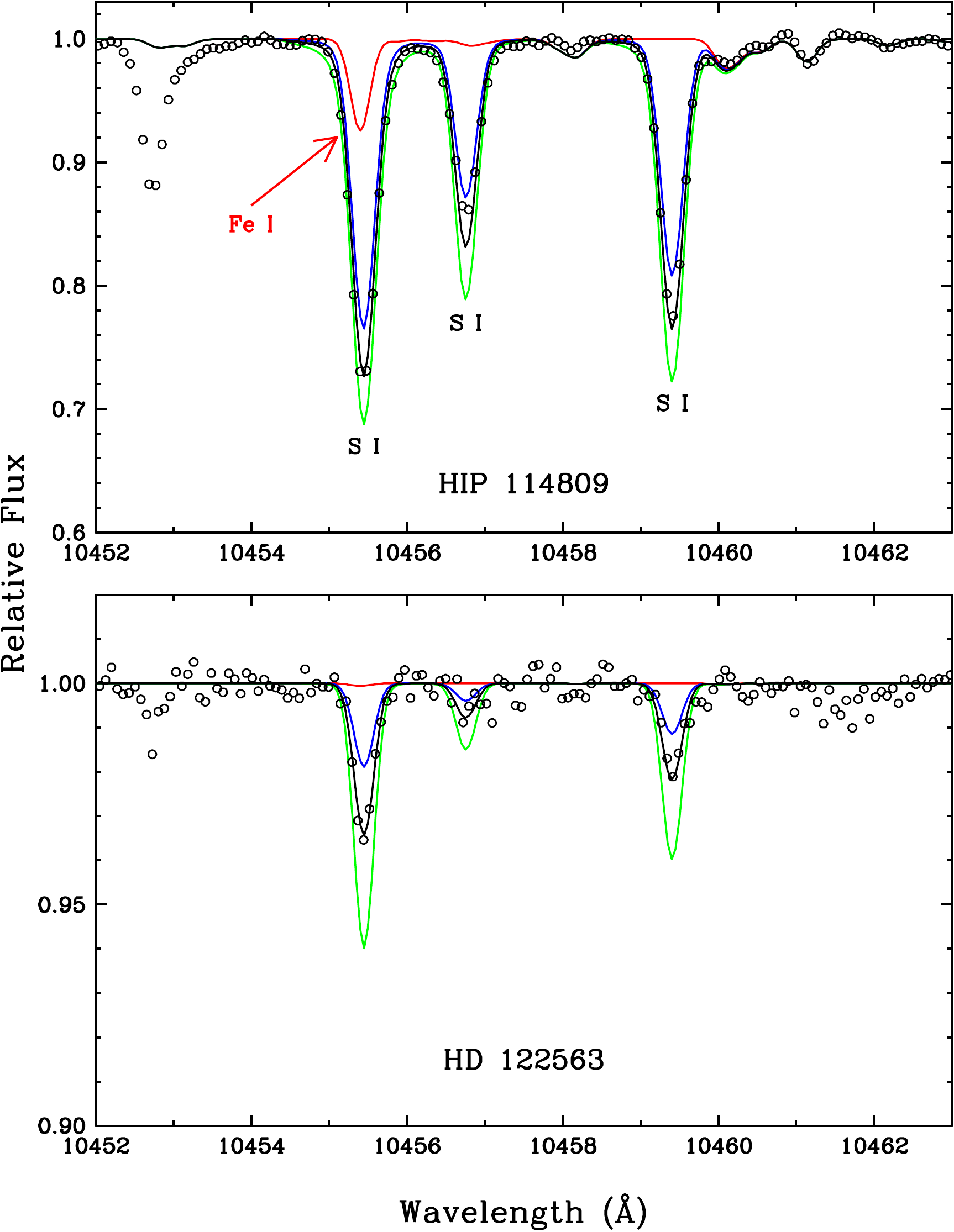}
\caption{\label{fig9}
\footnotesize
   \species{S}{i} in the RHB star HIP 114809 and in the very metal-poor giant
   HD 122563.
   In each panel the observed spectra are shown with open circles.
   The red line represents the synthetic spectrum with no contribution from
   \species{S}{i}, the black line is for the best overall S abundance, and the
   blue and green lines are for S abundances that are 0.3~dex smaller and larger
   than the best abundance. 
}
\end{figure}

We have detected \species{S}{i} lines in HD~122563, as we illustrate in
Figure~\ref{fig9}.
The triplet of lines near 10457~\AA\ is strong in metal-rich giants such
as HIP~114809 (top panel), and the 10455~\AA\ line is blended enough that
its abundance must be derived by synthetic spectrum computations.
The two strongest lines of the triplet are clearly seen in HD~122563
(bottom panel).
We have derived an S abundance from two of these lines 
and all three of the 9220~\AA\ lines.  
Our LTE abundance is [S/Fe] = +0.75 ($\sigma$ = 0.26), and if we apply
the non-LTE correction of $-$0.17 computed by \cite{spite11} we get
[S/Fe] = +0.58, somewhat larger than their value of +0.42.
Further investigation of this abundance with a completely self-consistent
models and computational techniques should be pursued in a future 
investigation.

\vspace*{0.2in}
\section{CONCLUSIONS}\label{discussion}

We have used the Habitable Zone Planet Finder to explore the spectra of red 
giant stars in the $zyJ$ (8100$-$12800~\AA) wavelength domain.
We studied HPF spectra of 13 RHB stars, one open cluster red clump star, and 
one very metal-poor halo giant.
We derived Fe metallicities that are consistent with those determined with
optical and H- \& K-band high-resolution spectra.
Abundance ratios also proved to be generally in agreement with results from
other spectral regions, but it is clear that many transitions in the HPF
domain need non-LTE computations to yield accurate results.

Our analyses highlighted four special giant stars.
HIP~33578 is an RHB with very low C/N and \carbiso\ ratios. 
It is possibly a member of the weak G-band subclass.
HIP~99789 is a (perhaps not so) rare Li-rich RHB star.
The unique HPF contribution is our discovery that this star has a strong
\species{He}{i} 10830~\AA\ absorption feature, perhaps indicative of 
unusually large atmospheric activity.
NGC~6940 MMU 105 is a normal open cluster red clump star; its HPF abundances
generally concur with those derived from other spectral regions.
Finally, the well-known very metal-poor giant HD~122563 has very few
observable features, as expected.
Notable among the detections is the 9220~\AA\ \species{S}{i} triplet, 
yielding an abundance in one of the lowest-metallicity cases.

HPF is dedicated to measuring high-precision radial velocities
in M-dwarf stars, but we have shown in this paper that it can be an attractive
instrument for investigating the chemical compositions of warmer red giants.
Most useful transitions arise from light and Fe-group neutral species.
For some elements, \eg, Mg, Si, and S, the abundances from the HPF $zyJ$ 
spectral region may prove to be superior to those from the optical region.

\acknowledgments

We thank Noriyuki Matsunaga and George Preston for helpful comments on the 
manuscript.
These results are based on observations obtained with the Habitable-zone 
Planet Finder Spectrograph on the Hobby-Eberly Telescope. 
We thank the Resident Astronomers and Telescope Operators at the HET for the 
skillful execution of our observations of our observations with HPF. 
The Hobby-Eberly Telescope is a joint project of the University of Texas at 
Austin, the Pennsylvania State University, Ludwig-Maximilians-Universität 
M{\" u}nchen, and Georg-August Universit{\" a}t Gottingen. 
The HET is named in honor of its principal benefactors, William P. Hobby and 
Robert E. Eberly. 
The HET collaboration acknowledges the support and resources from the Texas 
Advanced Computing Center.
We acknowledge support from NSF grants AST-1006676, AST-1126413, AST-1310885, 
AST-1517592, AST-1310875, AST-1910954, AST-1907622, AST-1909506, and support 
from the Heising-Simons foundation in our pursuit of precision spectroscopy 
in the NIR.

This work also used the Immersion Grating Infrared Spectrometer (IGRINS) that 
was developed under a collaboration between the University of Texas at Austin 
and the Korea Astronomy and Space Science Institute (KASI) with the financial 
support of the US National Science Foundation under grants AST-1229522 and 
AST-1702267, of the McDonald Observatory of the University of Texas at Austin, 
and of the Korean GMT Project of KASI.
Support for this study was also provided by National Science Foundation
grant AST-1616040.

\software{linemake (https://github.com/vmplacco/linemake), 
MOOG (Sneden 1973), 
IRAF (Tody 1986, Tody 1993), 
SPECTRE (Fitzpatrick \& Sneden 1987), 
Goldilocks (https://github.com/grzeimann/Goldilocks\_Documentation), 
PySME (Wehrhahn 2019; 10.5281/zenodo.3520617)}


\clearpage

\clearpage

\clearpage                                                    

\clearpage                                                    

\clearpage

\clearpage

\clearpage

\clearpage
\startlongtable
\begin{center}
\begin{deluxetable}{lccrrrrc}
\tabletypesize{\footnotesize}
\tablewidth{0pt}
\tablecaption{Basic Parameters of Target Stars and Observation Dates\label{tab-log}}
\tablecolumns{8}
\tablehead{
\colhead{Star Name}                               &
\colhead{RA(2000)\tablenotemark{\scriptsize a}}  &
\colhead{DEC(2000)\tablenotemark{\scriptsize a}} &
\colhead{B\tablenotemark{\scriptsize{a}}}         &
\colhead{V\tablenotemark{\scriptsize{a}}}         &
\colhead{K\tablenotemark{\scriptsize{a}}}         &
\colhead{d\tablenotemark{\scriptsize{b}}}         &
\colhead{Date observed}                           \\
\colhead{}                                        &
\colhead{(h m s)}                                 &
\colhead{($^\circ$ $\prime$  $\prime$$\prime$)}   &
\colhead{(mag)}                                   &
\colhead{(mag)}                                   &
\colhead{(mag)}                                   &
\colhead{(pc)}                                    &
\colhead{}                               
}
\startdata
\multicolumn{8}{c}{Observed Targets} \\
HIP 476         & 00 05 41.96 & +13 23 46.5 &  6.43 &  5.55 & 3.77 &  122 & 2018-10-11	\\
HIP 19740       & 04 13 56.38 & +09 15 49.7 &  5.71 &  4.89 & 2.97 &  128 & 2018-10-10	\\
HIP 27091       & 05 44 41.43 & +12 24 55.1 &  7.89 &  6.94 & 4.76 &  291 & 2018-11-04	\\
HIP 28417       & 06 00 06.04 & +27 16 19.8 &  7.64 &  6.62 & 4.21 &  145 & 2018-10-04	\\
HIP 29962       & 06 18 26.38 & +49 11 09.1 &  8.63 &  7.74 & 5.54 &  199 & 2018-10-09	\\
HIP 32431       & 06 46 10.37 & +23 22 16.4 &  7.34 &  6.49 & 4.32 &  221 & 2018-10-09	\\
HIP 33578       & 06 58 36.73 & $-$07 11 08.9 &  7.89 &  7.13 & 5.14 &  145 & 2018-11-19	\\
HIP 57748       & 11 50 36.25 & +43 39 32.2 &  8.66 &  7.90 & 6.01 &  210 & 2019-05-18	\\
HIP 99789       & 20 14 46.98 & +26 47 32.6 &  9.20 &  8.33 & 6.15 &  470 & 2018-11-21	\\
HIP 106775      & 21 37 41.66 & +15 12 57.5 &  8.37 &  7.44 & 5.24 &  358 & 2018-11-16	\\
HIP 113610      & 23 00 37.91 & +00 11 09.1 &  7.10 &  6.23 & 4.22 &  183 & 2018-10-08	\\
HIP 114809      & 23 15 23.04 & +25 40 20.1 &  7.64 &  6.81 & 4.76 &  209 & 2018-11-12	\\
HIP 116053      & 23 30 54.85 & +07 59 48.9 &  9.04 &  8.10 & 5.89 &  243 & 2018-11-19	\\
NGC6940 MMU 105  & 20 34 25.46 & +28 05 05.6 & 11.90 & 10.66 & 7.82 & 1021 & 2019-06-19	\\
HD 122563       & 14 02 31.84 & +09 41 09.9 &  7.10 &  6.19 & 3.73 &  290 & 2018-07-17	\\
\multicolumn{8}{c}{Telluric Standards} \\
HR 26	        & 00 10 02.20 &   +11 08 44.9 & 5.46 & 5.53 & 5.70 &  93 & 2018-10-11 \\
HR 1307	        & 04 13 34.56 &   +10 12 44.9 & 6.27 & 6.25 & 6.03 & 145 & 2018-10-10 \\
HR 1808        	& 05 27 10.09 &   +17 57 43.9 & 5.30 & 5.40 & 5.59 & 203 & 2018-11-04 \\
HR 2084        	& 05 57 59.65 &   +25 57 14.1 & 4.76 & 4.82 & 5.03 & 476 & 2018-10-04 \\
HR 2257        	& 06 22 03.55 &   +59 22 19.5 & 6.21 & 6.07 & 5.51 & 152 & 2018-10-09 \\
HR 2529        	& 06 51 33.05 &   +21 45 40.1 & 5.26 & 5.26 & 5.17 & 129 & 2018-10-09 \\
HR 2648        	& 07 02 54.77 &   $-$04 14 21.2 & 4.80 & 5.00 & 5.50 & 373 & 2018-11-19 \\
HIP 55182       & 11 17 55.15 &   +40 50 14.4 & 8.98 & 8.93 & 8.88 & 345 & 2019-05-18 \\
HR 7734	        & 20 14 04.88 &   +36 36 17.5 & 6.45 & 6.46 & 6.36 & 283 & 2018-11-21 \\
HR 8258	        & 21 35 27.03 &   +24 27 07.7 & 6.35 & 6.22 & 5.79 & 198 & 2018-11-16 \\
HR 8773	        & 23 03 52.61 &   +03 49 12.1 & 4.40 & 4.52 & 4.75 & 125 & 2018-10-08 \\
HIP 115579      & 23 24 43.38 &   +36 21 43.8 & 6.89 & 7.02 & 7.43 & 413 & 2018-11-12 \\
HR 8963	        & 23 37 56.80 &   +18 24 02.4 & 5.48 & 5.48 & 5.42 &  74 & 2018-11-19 \\
HR 7958	        & 20 46 38.59 &   +46 31 54.1 & 6.36 & 6.30 & 6.04 & 151 & 2019-06-19 \\
HIP 68209       & 13 57 52.12 &   +16 12 07.5 & 7.65 & 7.58 & 7.41 & 256 & 2018-07-17 \\
\enddata        

\tablenotetext{a}{SIMBAD Database}
\tablenotetext{b}{Adopted from Gaia DR2 \citep{Gaia18b}}

\end{deluxetable}                                                   
\end{center}

\begin{center}                                                  
\begin{deluxetable}{lrrrrrrrr}                                  
\tablewidth{0pt}                                                
\tablecaption{Radial Velocities\label{tab-vel}}                 
\tablecolumns{9}                                                
\tablehead{                                                     
\colhead{Star}                           &                      
\colhead{RV\tablenotemark{\scriptsize a}}            &                      
\colhead{$\sigma$}                       &                      
\colhead{RV}                             &                      
\colhead{$\sigma$}                       &                      
\colhead{RV}                             &                      
\colhead{$\sigma$}                       &                      
\colhead{$\delta$RV}                     &                      
\colhead{$\delta$RV}                     \\                     
\colhead{}                               &                      
\colhead{HPF}                            &                      
\colhead{HPF}                            &                      
\colhead{opt}                            &                      
\colhead{opt}                            &                      
\colhead{Gaia}                           &                      
\colhead{Gaia}                           &                      
\colhead{opt$-$HPF}                        &                    
\colhead{Gaia$-$HPF}                                            
}                                                               
\startdata                                                      
HIP 476     &     2.06 & 0.17 &     2.31 & 0.12 &     1.90 & 0.12 &    0.25 & $-$0.16 \\
HIP 19740   &  $-$7.69 & 0.23 &  $-$7.76 & 0.18 &  $-$7.98 & 0.14 & $-$0.07 & $-$0.30 \\
HIP 27091   & $-$37.98 & 0.26 & $-$37.13 & 0.18 & $-$38.18 & 0.15 &    0.85 & $-$0.20 \\
HIP 28417   &   100.59 & 0.18 &   101.21 & 0.21 &   100.22 & 0.18 &    0.62 & $-$0.36 \\
HIP 29962   &    11.99 & 0.14 &    15.89 & 0.21 &    12.13 & 1.38 &    3.90 &    0.14 \\                                                          
HIP 32431   &    41.18 & 0.15 &    41.08 & 0.20 &    41.15 & 0.17 & $-$0.10 & $-$0.03 \\                                                      
HIP 33578   &    24.33 & 0.13 &    23.04 & 0.17 &    24.12 & 0.20 & $-$1.29 & $-$0.21 \\                                                      
HIP 57748   & $-$20.48 & 0.16 & $-$21.04 & 0.16 & $-$20.88 & 0.13 & $-$0.56 & $-$0.40 \\                                                
HIP 99789   & $-$16.90 & 0.20 & $-$12.90 & 0.16 & $-$17.96 & 0.86 &    4.00 & $-$1.06 \\                                                  
HIP 106775  & $-$27.22 & 0.09 & $-$27.89 & 0.18 & $-$27.53 & 0.23 & $-$0.67 & $-$0.30 \\                                                
HIP 113610  & $-$14.48 & 0.07 & $-$13.75 & 0.15 & $-$14.53 & 0.14 &    0.73 & $-$0.05 \\                                                  
HIP 114809  &    38.13 & 0.17 &    41.84 & 0.19 &    38.06 & 0.18 &    3.71 & $-$0.07 \\                                                        
HIP 116053  &     7.09 & 0.24 &     5.05 & 0.52 &     7.13 & 0.18 & $-$2.03 &    0.04 \\                                                        
N6940 MMU 105  &     9.13 & 0.20 &     7.74 & 0.08 &     8.38 & 0.23 & $-$1.39 & $-$0.75 \\                                                      
HD 122563   & $-$25.85 & 0.26 & $-$27.09 & 0.56 & $-$26.13 & 0.04 & $-$1.24 & $-$0.28 \\                                                
\enddata
                                                                
\tablenotetext{a}{All velocities are in \kmsec}
                                                                
\end{deluxetable}                                               
\end{center}

\begin{center}
\begin{deluxetable}{lrrrrrrrrrrrrr}
\tabletypesize{\footnotesize}
\tablewidth{0pt}
\tablecaption{Program Star Model Atmospheres\label{tab-models}}
\tablecolumns{14}
\tablehead{
\colhead{Star Name}                      &
\colhead{\teff}                          &
\colhead{\logg}                          &
\colhead{\vmicro}                        &
\colhead{[\species{Fe}{i}]}              &
\colhead{$\sigma$}                       &
\colhead{[\species{Fe}{ii}]}             &
\colhead{$\sigma$}                       &
\colhead{[\species{Fe}{i}]}              &
\colhead{$\sigma$}                       &
\colhead{\#lines}                        &
\colhead{[\species{Fe}{ii}]}             &
\colhead{$\sigma$}                       &
\colhead{\#lines}                        \\
\colhead{}                               &
\colhead{K}                              &
\colhead{}                               &
\colhead{\kmsec}                         &
\colhead{}                               &
\colhead{}                               &
\colhead{}                               &
\colhead{}                               &
\colhead{}                               &
\colhead{}                               &
\colhead{}                               &
\colhead{}                               &
\colhead{}                               &
\colhead{}                               \\
\colhead{}                               &
\colhead{opt\tablenotemark{\scriptsize a}}           &
\colhead{opt}                            &
\colhead{opt}                            &
\colhead{opt}                            &
\colhead{opt}                            &
\colhead{opt}                            &
\colhead{opt}                            &
\colhead{HPF}                            &
\colhead{HPF}                            &
\colhead{HPF}                            &
\colhead{HPF}                            &
\colhead{HPF}                            &
\colhead{HPF}                            
}
\startdata
\multicolumn{10}{c}{Field RHB Stars} \\
HIP 476      & 5109 & 2.68 & 1.36 & $-$0.10 &  0.07 & $-$0.09 & 0.04 & $-$0.02  & 0.14  & 21 &  $-$0.03 & 0.04 & 4 \\
HIP 19740    & 5135 & 2.51 & 1.35 & $-$0.10 &  0.07 & $-$0.15 & 0.03 & $-$0.09  & 0.11  & 22 &  $-$0.12 & 0.03 & 4 \\
HIP 27091    & 5001 & 2.37 & 1.56 & $-$0.21 &  0.06 & $-$0.24 & 0.07 & $-$0.14  & 0.11  & 19 &  $-$0.16 & 0.09 & 4 \\
HIP 28417    & 4679 & 2.15 & 1.42 & $-$0.38 &  0.06 & $-$0.43 & 0.05 & $-$0.33  & 0.12  & 20 &  $-$0.27 & 0.12 & 3 \\
HIP 29962    & 5118 & 2.92 & 1.19 & $-$0.28 &  0.06 & $-$0.36 & 0.04 & $-$0.23  & 0.11  & 20 &  $-$0.21 & 0.08 & 4 \\
HIP 32431    & 5125 & 2.26 & 1.57 & $-$0.56 &  0.05 & $-$0.63 & 0.05 & $-$0.51  & 0.11  & 25 &  $-$0.53 & 0.03 & 4 \\
HIP 33578    & 5118 & 2.92 & 1.19 & $-$0.24 &  0.06 & $-$0.28 & 0.05 & $-$0.30  & 0.15  & 22 &  $-$0.21 & 0.01 & 3 \\
HIP 57748    & 5307 & 2.34 & 1.82 & $-$0.17 &  0.06 & $-$0.19 & 0.05 & $-$0.17  & 0.12  & 22 &  $-$0.13 & 0.05 & 4 \\
HIP 99789    & 5054 & 2.41 & 1.12 & $-$0.22 &  0.06 & $-$0.28 & 0.09 & $-$0.06  & 0.14  & 20 &  $-$0.29 & 0.02 & 4 \\
HIP 106775   & 5111 & 2.41 & 1.61 & $-$0.21 &  0.05 & $-$0.26 & 0.06 & $-$0.15  & 0.14  & 20 &  $-$0.15 & 0.07 & 4 \\
HIP 113610   & 5106 & 2.39 & 1.55 & $-$0.07 &  0.07 & $-$0.04 & 0.05 & $-$0.10  & 0.13  & 19 & \nodata & \nodata & \nodata \\
HIP 114809   & 5139 & 2.59 & 1.31 & $-$0.33 &  0.06 & $-$0.38 & 0.05 & $-$0.32  & 0.10  & 23 &  $-$0.31 & 0.04 & 3 \\
HIP 116053   & 5100 & 3.09 & 0.98 & $-$0.07 &  0.06 & $-$0.08 & 0.05 &    0.03  & 0.13  & 24 &  $-$0.01 & 0.08 & 3 \\
\multicolumn{10}{c}{Other Stars} \\
N6940 MMU 105& 4765 & 2.34 & 1.35 & $-$0.10 &  0.08 & $-$0.15 & 0.07 & $-$0.04  & 0.11  & 16 &     0.02 & 0.03 & 4 \\
  HD122563   & 4500 & 0.80 & 2.20 & $-$2.98 &  0.07 & $-$2.98 & 0.07 & $-$2.89  & 0.03  &  7 & \nodata & \nodata & \nodata \\
\enddata                                                            

\tablenotetext{a}{The optical values are taken from high-resolution optical 
                  spectroscopy obtained with the McDonald Observatory 2.7m 
                  TS23 echelle spectrograph 
                  (Af\c{s}ar16, BT16, Af\c{s}ar18a).}

\end{deluxetable}                                                   
\end{center}

\begin{center}
\begin{deluxetable}{cccrl}
\tablewidth{0pt}
\tablecaption{Atomic Lines\tablenotemark{\scriptsize a}\label{tab-lines}}
\tablecolumns{5}
\tablehead{
\colhead{$\lambda$}                      &
\colhead{Species}                        &
\colhead{$\chi$}                         &
\colhead{log($gf$)}                      &
\colhead{Source\tablenotemark{\scriptsize b}}        \\
\colhead{(\AA)}                          &
\colhead{}                               &
\colhead{(eV)}                           &
\colhead{}                               &
\colhead{}                               
}
\startdata
    8335.15 & \species{C}{i} &  7.679 &  $-$0.44 &  NISTB+ \\
    9061.44 & \species{C}{i} &  7.477 &  $-$0.35 &  NISTB  \\
    9094.83 & \species{C}{i} &  7.482 &  $+$0.15 &  NISTB  \\
    9405.73 & \species{C}{i} &  7.679 &  $+$0.29 &  NISTB+ \\
    9658.43 & \species{C}{i} &  7.482 &  $-$0.28 &  NISTB  \\
   10123.87 & \species{C}{i} &  8.531 &  $-$0.03 &  NISTC+ \\
   10683.08 & \species{C}{i} &  7.477 &  $+$0.08 &  NISTD  \\
   10685.34 & \species{C}{i} &  7.475 &  $-$0.27 &  NISTB  \\
   10691.24 & \species{C}{i} &  7.482 &  $+$0.34 &  NISTB  \\
   10707.32 & \species{C}{i} &  7.477 &  $-$0.41 &  NISTB  \\
\enddata

\tablenotetext{a}{The full version of this table in ascii form is available
                 on$-$line}

\tablenotetext{b}{Sources of the transition probabilities:  NIST =
                  the NIST Atomic Spectra Database, 
                  https://physics.nist.gov/asd, with notation of
                  their assessment of log($gf$) quality; LAWLER13 =
                  \cite{lawler13}; KURUCZ = the \cite{kurucz11,kurucz18}
                  compendium, http://kurucz.harvard.edu/linelists.html;
                  DENHARTOG14 = \cite{denhartog14}; RUFFONI14 =
                  \cite{ruffoni14}; OBRIAN91 = \cite{obrian91};
                  LAWLER15 = \cite{lawler15}. }

\end{deluxetable}
\end{center}

\begin{center}                                                
\begin{deluxetable}{ccccccccccccccccc}                                    
\tablewidth{0pt}                                              
\tablecaption{Line Abundances\tablenotemark{\scriptsize a}\label{tab-lineabs}}                  
\tablecolumns{17}                                              
\tabletypesize{\scriptsize}
\tablehead{
\colhead{Species}                        &
\colhead{$\lambda$}                      &                    
\colhead{(1)\tablenotemark{\scriptsize b}}           &                    
\colhead{(2)}                            &                    
\colhead{(3)}                            &                    
\colhead{(4)}                            &                    
\colhead{(5)}                            &                    
\colhead{(6)}                            &                        
\colhead{(7)}                            &                    
\colhead{(8)}                            &                    
\colhead{(9)}                            &                    
\colhead{(10)}                           &                    
\colhead{(11)}                           &                    
\colhead{(12)}                           &                    
\colhead{(13)}                           &                    
\colhead{(14)}                           &                    
\colhead{(15)}                           \\                   
\colhead{}                                 &                    
\colhead{(\AA)}                          &
\colhead{m\AA}                           &
\colhead{m\AA}                           &
\colhead{m\AA}                           &
\colhead{m\AA}                           &
\colhead{m\AA}                           &
\colhead{m\AA}                           &
\colhead{m\AA}                           &
\colhead{m\AA}                           &
\colhead{m\AA}                           &
\colhead{m\AA}                           &
\colhead{m\AA}                           &
\colhead{m\AA}                           &
\colhead{m\AA}                           &
\colhead{m\AA}                           &
\colhead{m\AA}                           
}                                                             
\startdata
\species{C}{i} &  8335.15 &    8.18 &    8.02 &    8.06 &    8.21 &    8.04 &    7.75 &    7.67 &    8.43 &    7.94 &    8.13 &    8.15 &    8.08 &    8.17 & \nodata & \nodata \\
\species{C}{i} &  9061.44 &    8.25 &    7.99 &    8.30 &    8.21 &    8.11 &    7.96 &    7.64 &    8.57 &    8.18 &    8.26 &    8.22 &    8.19 &    8.29 & \nodata & \nodata \\
\species{C}{i} &  9094.83 &    8.29 &    8.09 &    8.34 &    8.16 &    8.16 &    8.02 &    7.68 &    8.64 &    8.24 &    8.27 &    8.30 &    8.24 &    8.19 & \nodata & \nodata \\
\species{C}{i} &  9405.73 &    8.17 & \nodata & \nodata &    8.31 & \nodata &    7.87 &    7.69 &    8.54 & \nodata & \nodata & \nodata &    8.16 &    8.12 & \nodata & \nodata \\
\species{C}{i} &  9658.43 &    8.29 &    8.08 &    8.25 & \nodata &    7.99 &    7.89 &    7.66 &    8.65 &    8.11 &    8.20 &    8.22 &    8.19 &    8.29 & \nodata & \nodata \\
\species{C}{i} & 10123.87 &    8.26 &    8.04 &    8.26 &    8.30 &    8.11 &    7.86 &    7.55 & \nodata & \nodata &    8.20 &    8.15 &    8.06 &    8.20 & \nodata & \nodata \\
\species{C}{i} & 10683.08 &    8.21 &    8.10 &    8.30 &    8.11 &    8.07 &    7.98 &    7.67 &    8.56 &    8.06 &    8.21 &    8.21 &    8.16 &    8.13 & \nodata & \nodata \\
\species{C}{i} & 10685.34 &    8.19 &    8.09 &    8.27 &    8.12 &    8.05 &    7.89 &    7.62 &    8.54 &    8.04 &    8.22 &    8.22 &    8.13 &    8.11 & \nodata & \nodata \\
\species{C}{i} & 10691.24 &    8.21 &    8.10 &    8.27 &    8.13 &    8.02 &    7.99 &    7.66 &    8.56 &    8.07 &    8.22 &    8.24 &    8.15 &    8.10 & \nodata & \nodata \\
\species{C}{i} & 10707.32 &    8.18 &    8.04 &    8.25 &    8.16 &    8.02 &    7.89 &    7.62 &    8.55 &    8.02 &    8.19 &    8.15 &    8.11 &    8.16 & \nodata & \nodata \\
\enddata                                                      

\tablenotetext{a}{The full version of this table in ascii form is available 
                 on$-$line}

\tablenotetext{b}{Star identifications:  (1) HIP~476; (2) HIP~19740,
               (3) HIP~27091; (4) HIP~28417; (5) HIP~29962: (6) HIP~32431;
               (7) HIP~33578; (8) HIP~57748; (9) HIP~99789;
               (10) HIP~106775; (11) HIP~113610; (12) HIP~114809;
               (13) HIP~116043; (14) NGC 6940 MMU 105; (15) HD122563}

\end{deluxetable}
\end{center}

\begin{center}                                                
\begin{deluxetable}{lrrrrrrrrrrrrrrr}                                    
\tablewidth{0pt}                                              
\tablecaption{Mean Abundances\label{tab-meanabs}}                
\tablecolumns{16}                                              
\tablehead{                                                   
\colhead{Star}                           &                    
\colhead{\species{C}{i}\tablenotemark{\scriptsize a}}&                    
\colhead{\species{Na}{i}}                &                    
\colhead{\species{Mg}{i}}                &                    
\colhead{\species{Si}{i}}                &                    
\colhead{\species{P}{i}}                 &                    
\colhead{\species{S}{i}}                 &                    
\colhead{\species{K}{i}}                 &                    
\colhead{\species{Ca}{i}}                &                    
\colhead{\species{Ti}{i}}                &                    
\colhead{\species{Ti}{ii}}               &                    
\colhead{\species{Cr}{i}}                &                    
\colhead{\species{Mn}{i}}                &                    
\colhead{\species{Co}{i}}                &                    
\colhead{\species{Ni}{i}}                &                    
\colhead{\species{Sr}{i}}                
}                                                             
\startdata                                                    
HIP 476\tablenotemark{\scriptsize b}
     &  $-$0.22&   0.18&   0.03&  $-$0.14&   0.02&   0.14&   0.35&   0.15&  $-$0.04&   0.00&  $-$0.01&  $-$0.01&  $-$0.11&   0.08&   0.38\\
       &   0.07&   0.07&   0.14&   0.09&   0.03&   0.04&   0.29&   0.10&   0.09&  $-$1.00&   0.13&  $-$1.00&   0.00&   0.14&   0.05\\
       &     16&      4&     11&     12&      3&      4&      3&      2&      6&      1&     13&      1&      2&      8&      3\\
HIP 19740&  $-$0.30&   0.17&   0.02&  $-$0.10&  $-$0.07&   0.09&   0.32&   0.12&  $-$0.03&   0.02&  $-$0.05& \nodata&  $-$0.51&   0.05&   0.50\\
       &   0.06&   0.09&   0.15&   0.10&   0.08&   0.08&   0.16&   0.11&   0.15&  $-$1.00&   0.16&  \nodata&   0.07&   0.10&   0.09\\
       &     15&      4&     10&     12&      3&      4&      3&      2&      6&      1&     13&      0&      2&      7&      3\\
HIP 27091&  $-$0.09&   0.30&   0.16&   0.01&   0.05&   0.21&   0.40&   0.26&   0.05& 994.19&  $-$0.03& \nodata&  $-$0.20&   0.19&   0.37\\
       &   0.12&   0.11&   0.07&   0.11&   0.00&   0.12&   0.21&   0.02&   0.10&  \nodata&   0.14&  \nodata&  $-$1.00&   0.07&   0.07\\
       &     15&      4&      8&     11&      3&      4&      2&      2&      6&      0&      8&      0&      1&      6&      3\\
HIP 28417&   0.09&   0.17&   0.20&   0.03&   0.21&   0.25&   0.31&   0.28&  $-$0.01&   0.08&  $-$0.04& \nodata&  $-$0.02&   0.18&   0.26\\
       &   0.08&   0.08&   0.05&   0.10&   0.05&   0.05&   0.13&   0.05&   0.04&  $-$1.00&   0.10&  \nodata&   0.11&   0.09&   0.08\\
       &     15&      4&      8&     12&      3&      3&      3&      2&      7&      1&     11&      0&      2&      7&      3\\
HIP 29962&  $-$0.17&   0.24&   0.11&  $-$0.05&   0.13&   0.17&   0.49&   0.23&   0.00&   0.04&  $-$0.05& \nodata&   0.06&   0.08&   0.37\\
       &   0.07&   0.10&   0.09&   0.07&   0.08&   0.02&   0.24&   0.11&   0.08&  $-$1.00&   0.12&  \nodata&   0.07&   0.12&   0.06\\
       &     15&      4&     13&     13&      3&      4&      3&      2&      7&      1&     10&      0&      2&      8&      3\\
HIP 32431&  $-$0.02&   0.20&   0.16&   0.09&   0.16&   0.16&   0.43&   0.17&   0.00& \nodata&  $-$0.05& \nodata&  $-$0.04&   0.03&   0.45\\
       &   0.07&   0.06&   0.11&   0.13&   0.05&   0.11&   0.20&   0.02&   0.11&  \nodata&   0.13&  \nodata&   0.00&   0.18&   0.10\\
       &     15&      4&     11&     13&      3&      4&      3&      2&      7&      0&      8&      0&      2&      6&      3\\
HIP 33578&  $-$0.50&   0.23&   0.19&   0.03&   0.20&   0.21&   0.61&   0.36&   0.00& \nodata&  $-$0.05&   0.03&   0.08&   0.14&   0.35\\
       &   0.05&   0.05&   0.09&   0.10&   0.08&   0.03&   0.32&   0.14&   0.05&  \nodata&   0.08&  $-$1.00&   0.07&   0.11&   0.07\\
       &     16&      4&     14&     14&      3&      4&      3&      2&      7&      0&      8&      1&      2&      8&      3\\
HIP 57748&   0.25&   0.03&   0.12&   0.13&   0.17&   0.36&   0.18&   0.12&  $-$0.15&  $-$0.03&   0.02& \nodata&   0.01&   0.12&   0.40\\
       &   0.09&   0.12&   0.10&   0.08&   0.08&   0.15&   0.10&   0.04&   0.12&  $-$1.00&   0.10&  \nodata&   0.00&   0.10&   0.05\\
       &     15&      4&     12&     12&      3&      4&      3&      2&      2&      1&      8&      0&      2&      7&      2\\
HIP 99789&  $-$0.30&   0.11&   0.00&  $-$0.11&  $-$0.11&   0.04&   0.36&   0.22&  $-$0.08&  $-$0.07&  $-$0.09& \nodata&  $-$0.06&   0.01&   0.36\\
       &   0.08&   0.09&   0.08&   0.07&   0.18&   0.04&   0.25&   0.11&   0.11&  $-$1.00&   0.13&  \nodata&  $-$1.00&   0.15&   0.11\\
       &     14&      4&     10&     10&      3&      4&      3&      2&      6&      1&     10&      0&      1&      7&      3\\
HIP 106775&  $-$0.10&   0.44&   0.13&   0.03&   0.20&   0.24&   0.40&   0.24&   0.03&   0.05&   0.01& \nodata&  $-$0.02&   0.14&   0.40\\
       &   0.07&   0.12&   0.08&   0.13&   0.03&   0.07&   0.07&   0.07&   0.08&  $-$1.00&   0.12&  \nodata&  $-$1.00&   0.12&   0.09\\
       &     15&      4&      9&     11&      3&      4&      3&      2&      5&      1&     11&      0&      1&      9&      3\\
HIP 113610&  $-$0.17&   0.41&   0.09&   0.04&   0.13&   0.31&   0.31&   0.18&  $-$0.05& \nodata&  $-$0.04& \nodata&  $-$0.04&   0.09&   0.42\\
       &   0.07&   0.12&   0.09&   0.14&   0.13&   0.08&   0.07&  $-$1.00&   0.08&  \nodata&   0.10&  \nodata&   0.00&   0.06&   0.00\\
       &     15&      4&      9&     11&      3&      4&      2&      1&      4&      0&      7&      0&      2&      5&      2\\
HIP 114809&  $-$0.00&   0.29&   0.16&   0.01&   0.21&   0.26&   0.49&   0.24&   0.01& \nodata&  $-$0.07&   0.05&  $-$0.06&   0.09&   0.49\\
       &   0.07&   0.07&   0.08&   0.11&   0.06&   0.09&   0.22&   0.14&   0.17&  \nodata&   0.08&  $-$1.00&   0.00&   0.12&   0.07\\
       &     16&      4&     10&     10&      3&      4&      3&      2&      6&      0&     11&      1&      2&      8&      3\\
HIP 116053&  $-$0.31&  $-$0.02&  $-$0.11&  $-$0.23&  $-$0.08&   0.08&   0.26&   0.15&  $-$0.09& \nodata&  $-$0.06& \nodata&  $-$0.14&   0.07&   0.38\\
       &   0.07&   0.10&   0.09&   0.10&   0.13&   0.04&   0.10&   0.20&   0.06&  \nodata&   0.13&  \nodata&   0.04&   0.09&   0.10\\
       &     16&      4&     11&     12&      3&      4&      3&      2&      7&      0&     12&      0&      2&      8&      3\\
NGC 6940 MMU 105&  $-$0.15&   0.27&   0.06&  $-$0.06&   0.01&   0.20&   0.35&   0.26&  $-$0.07&   0.02&   0.03& \nodata&  $-$0.07&   0.18&   0.36\\
       &   0.09&   0.10&   0.14&   0.11&   0.00&   0.05&   0.00&   0.11&   0.12&  $-$1.00&   0.14&  \nodata&  $-$1.00&   0.11&   0.07\\
       &     12&      4&      8&     12&      2&      4&      2&      2&      5&      1&     10&      0&      1&      4&      3\\
\enddata  

\tablenotetext{a}{The \species{C}{i} mean abundances are tabulated here, 
                  but discussion of carbon is referred to \ref{cno}}

\tablenotetext{b}{For each star there are three rows, giving in order the 
                  mean abundances, the standard deviations $\sigma$, and the 
                  number of lines contributing to the means.}

\end{deluxetable}
\end{center}

\clearpage
\begin{center}
\begin{deluxetable}{lrrrrrr}
\tablewidth{0pt}
\tablecaption{Mean Non-LTE Abundance Corrections\label{tab-nltecorr}}
\tablecolumns{7}
\tablehead{
\colhead{Species}                                        &
\colhead{$\langle\Delta_{corr}\rangle$\tablenotemark{\scriptsize a}} &
\colhead{$\sigma$}                                       &
\colhead{\#lines}                                        &
\colhead{$\langle\Delta_{corr}\rangle$\tablenotemark{\scriptsize a}} &
\colhead{$\sigma$}                                       &
\colhead{\#lines}                                        \\ 
\colhead{}                                               &
\colhead{RHB}                                            &
\colhead{RHB}                                            &
\colhead{RHB}                                            &
\colhead{HD122\tablenotemark{\scriptsize b}}             &
\colhead{HD122\tablenotemark{\scriptsize b}}             &
\colhead{HD122\tablenotemark{\scriptsize b}}         
}
\startdata
\species{C}{i}   & $-$0.15  & 0.06  & 16  & $-$0.13  &    0.00  &       2 \\
\species{Na}{i}  & $-$0.16  & 0.19  &  5  & \nodata  & \nodata  & \nodata \\
\species{Mg}{i}  & $-$0.09  & 0.08  & 11  & $-$0.67  & \nodata  &       1 \\
\species{Si}{i}  & $-$0.19  & 0.11  & 14  & \nodata  & \nodata  & \nodata \\
\species{K}{i}   & $-$0.23  & 0.15  & 13  & \nodata  & \nodata  & \nodata \\
\species{Ca}{i}  & $-$0.12  & 0.10  &  2  & $-$0.25  &    0.07  &       3 \\
\enddata

\tablenotetext{b}{$\Delta_{corr}$ is the computed shift for a line need
                  to correct its LTE abundance for non-LTE effects}
\tablenotetext{b}{HD122 = HD 122563}

\end{deluxetable}
\end{center}

\begin{center}
\begin{deluxetable}{lrrcrrr}
\tablewidth{0pt}
\tablecaption{LiCNO Abundances\label{tab-licno}}
\tablecolumns{7}
\tablehead{
\colhead{Star}                                      &
\colhead{log $\epsilon$(Li)}                        &
\colhead{[C$_{C I}$/Fe]}                            &
\colhead{[C$_{mean}$/Fe]\tablenotemark{\scriptsize a}}          &
\colhead{[N$_{CN}$/Fe]\tablenotemark{\scriptsize b}}            &
\colhead{[O$_{O I}$/Fe]}                            &
\colhead{\carbiso}
}
\startdata
\multicolumn{7}{c}{optical spectra} \\
HIP 476        &       0.72  &              $-$0.35  &      $-$0.37  &             0.62  &      $-$0.05  &          \nodata \\
HIP 19740      &      $<$0   &              $-$0.47  &      $-$0.54  &             0.66  &      $-$0.11  &               17 \\
HIP 27091      &       1.28  &              $-$0.25  &      $-$0.42  &             0.73  &         0.02  &               20 \\
HIP 28417      &      $<$0   &              $-$0.02  &      $-$0.25  &             0.33  &         0.07  &               12 \\
HIP 29962      &      $<$0   &              $-$0.23  &      $-$0.36  &             0.61  &         0.12  &               20 \\
HIP 32431      &      $<$0   &              $-$0.12  &      $-$0.05  &             0.58  &         0.31  &               20 \\
HIP 33578      &      $<$0   &              $-$0.72  &      $-$0.88  &             0.67  &         0.17  &                3 \\
HIP 57748      &      $<$0   &              $-$0.04  &      $-$0.24  &             0.40  &      $-$0.07  &                6 \\
HIP 99789      &       2.61  &              $-$0.37  &      $-$0.51  &             0.66  &      $-$0.17  &               13 \\
HIP 106775     &      $<$0   &              $-$0.22  &      $-$0.43  &             0.76  &         0.07  &               13 \\
HIP 113610     &      $<$0   &              $-$0.46  &      $-$0.53  &             0.53  &      $-$0.12  &               20 \\
HIP 114809     &      $<$0   &              $-$0.24  &      $-$0.35  &             0.57  &       0.11  &            \nodata \\
HIP 116053     &      0.42   &              $-$0.21  &      $-$0.30  &             0.34  &      $-$0.01  &               17 \\
N6940 MMU 105  &      $<$0   &              $-$0.27  &      $-$0.28  &             0.41  &      $-$0.14  &               15 \\
\multicolumn{7}{c}{HPF spectra} \\
HIP 476        &   \nodata   &  $-$0.22$\pm$0.07   &   \nodata   &   0.54$\pm$0.05   &   \nodata   &           $>$20  \\
HIP 19740      &   \nodata   &  $-$0.30$\pm$0.06   &   \nodata   &   0.70$\pm$0.08   &   \nodata   &         \nodata  \\
HIP 27091      &   \nodata   &  $-$0.09$\pm$0.12   &   \nodata   &   0.58$\pm$0.08   &   \nodata   &  15$^{+5}_{-3}$  \\
HIP 28417      &   \nodata   &     0.09$\pm$0.08   &   \nodata   &   0.37$\pm$0.08   &   \nodata   &  10$^{+2}_{-2}$  \\
HIP 29962      &   \nodata   &  $-$0.17$\pm$0.07   &   \nodata   &   0.61$\pm$0.05   &   \nodata   &         \nodata  \\
HIP 32431      &   \nodata   &  $-$0.02$\pm$0.07   &   \nodata   &   0.58$\pm$0.05   &   \nodata   &  10$^{+5}_{-3}$  \\
HIP 33578      &   \nodata   &  $-$0.50$\pm$0.05   &   \nodata   &   0.68$\pm$0.05   &   \nodata   &   3$^{+2}_{-2}$  \\
HIP 57748      &   \nodata   &     0.25$\pm$0.09   &   \nodata   &   0.37$\pm$0.05   &   \nodata   &         \nodata  \\
HIP 99789      &   \nodata   &  $-$0.30$\pm$0.08   &   \nodata   &   0.52$\pm$0.05   &   \nodata   &         \nodata  \\
HIP 106775     &   \nodata   &  $-$0.10$\pm$0.07   &   \nodata   &   0.70$\pm$0.05   &   \nodata   &   9$^{+4}_{-2}$  \\
HIP 113610     &   \nodata   &  $-$0.17$\pm$0.07   &   \nodata   &   0.61$\pm$0.10   &   \nodata   &         \nodata  \\
HIP 114809     &   \nodata   &     0.00$\pm$0.07   &   \nodata   &   0.56$\pm$0.05   &   \nodata   &  10$^{+10}_{-3}$ \\
HIP 116053     &   \nodata   &  $-$0.31$\pm$0.07   &   \nodata   &   0.24$\pm$0.08   &   \nodata   &  25$^{+10}_{-5}$ \\
N6940 MMU 105  &   \nodata   &  $-$0.15$\pm$0.09   &   \nodata   &   0.39$\pm$0.05   &   \nodata   &  15$^{+5}_{-2}$  \\
\enddata

\tablenotetext{a}{The optical mean C abundances are based on CH and C$_2$ 
                  features.}                   
                                                              
\tablenotetext{b}{Both optical and HPF N abundances assume the C and O      
                  abundances determined from optical [\species{O}{i}]      
                  transitions.}                               

\end{deluxetable}                                                               \end{center}

\clearpage
\begin{center}
\begin{deluxetable}{lcrrrc}
\tablewidth{0pt}
\tablecaption{HD~122563 Line Abundances\tablenotemark{\scriptsize a}\label{tab-hd122lines}}
\tablecolumns{6}
\tablehead{
\colhead{Wavelength}                     &
\colhead{Species}                        &
\colhead{$\chi$}                         &
\colhead{log($gf$)}                      &
\colhead{log $\epsilon$}                 &
\colhead{Source\tablenotemark{\scriptsize b}}        \\
\colhead{\AA}                            &
\colhead{}                               &
\colhead{eV}                             &
\colhead{}                               &
\colhead{}                               &
\colhead{}                                
}
\startdata
10683.08  &  \species{C}{i} & 7.477  &      0.08  &  5.38  &   NIST D  \\
10691.24  &  \species{C}{i} & 7.482  &      0.35  &  5.53  &   NIST B  \\
 8806.76  & \species{Mg}{i} & 4.343  &   $-$0.14  &  5.29  &   NIST A  \\
10288.94  & \species{Si}{i} & 4.916  &   $-$1.48  &  4.96  &   NIST C+ \\
10371.26  & \species{Si}{i} & 4.926  &   $-$0.71  &  4.94  &   NIST C+ \\
10660.91  & \species{Si}{i} & 4.916  &   $-$0.32  &  5.22  &   NIST C+ \\
10689.72  & \species{Si}{i} & 5.949  &      0.09  &  5.35  &   NIST C+ \\
10694.25  & \species{Si}{i} & 5.959  &      0.26  &  5.37  &   NIST C+ \\
10727.41  & \species{Si}{i} & 5.979  &      0.42  &  5.41  &   NIST B  \\
10749.38  & \species{Si}{i} & 4.926  &   $-$0.27  &  5.23  &   NIST C+ \\ 
\enddata

\tablenotetext{a}{The full version of this table in ascii form is available
                 on$-$line}

\tablenotetext{b}{As defined in Table~\ref{tab-lineabs}}

\end{deluxetable}
\end{center}

\begin{center}
\begin{deluxetable}{lcrr}
\tablewidth{0pt}
\tablecaption{HD~122563 Mean Abundances\label{tab-hd122means}}
\tablecolumns{4}
\tablehead{
\colhead{Species}                        &
\colhead{$\langle$[X/Fe]$\rangle$}       &
\colhead{$\sigma$}                       &
\colhead{\#lines}                         
}
\startdata
\species{C}{i}   & $-$0.11 &    0.11 &  2 \\
\species{Mg}{i}  &    0.56 & \nodata &  1 \\
\species{Si}{i}  &    0.49 &    0.09 & 10 \\
\species{S}{i}   &    0.75 &    0.26 &  5 \\
\species{Ca}{i}  &    0.40 & \nodata &  1 \\
\species{Ca}{ii} &    0.54 &    0.02 &  2 \\
\species{Ti}{i}  &    0.23 &    0.12 &  3 \\
\species{Fe}{i}  & $-$2.87\tablenotemark{\scriptsize a} &    0.07 & 10 \\
\species{Sr}{ii} &    0.45 &    0.14 &  3 \\
\enddata

\tablenotetext{a}{For \species{Fe}{i} we quote the metallicity [Fe/H].}

\end{deluxetable}
\end{center}

\end{document}